\documentclass[12pt,a4paper,titlepage]{report}
\usepackage{epsfig} 
\usepackage{amssymb} 
\usepackage{amsmath}
\usepackage{graphicx}
\usepackage{caption}
\usepackage{subcaption}
\usepackage{natbib}
\usepackage{titling}
\bibpunct{(}{)}{;}{a}{}{,}
\title{Estimating the Sunyaev-Zel'dovich Signal from Quasar Hosts using a Halo Occupation Distribution based approach}
\author{Dhruba Dutta Chowdhury \\ Department of Physics, Presidency University\\\\Thesis Adviser\\Dr. Suchetana Chatterjee\\Assistant Professor, Department of Physics, Presidency University\\}
\date{}

\pretitle{
\begin{center}
\LARGE
\includegraphics[scale=0.5]{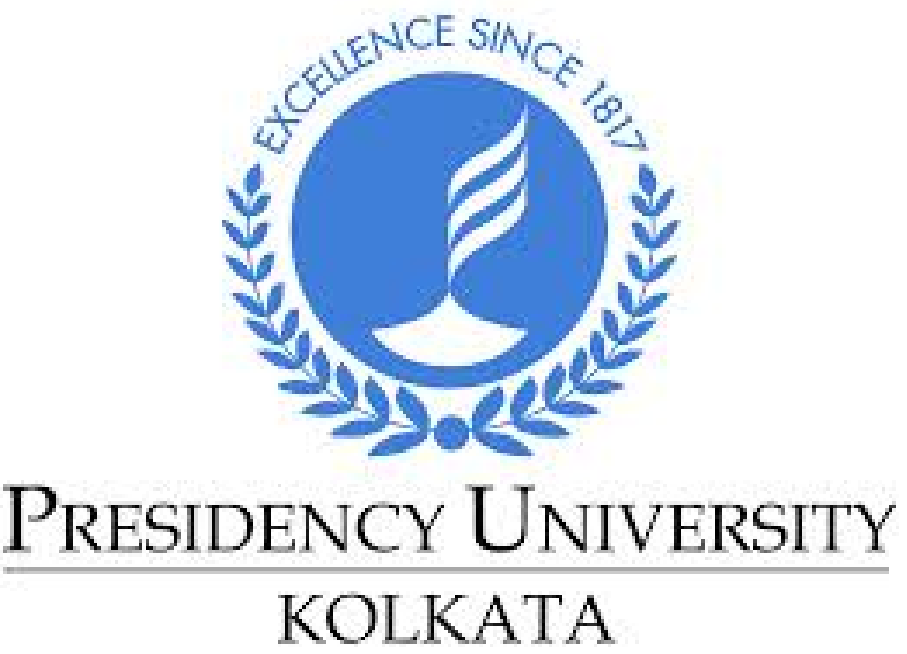}\\[\bigskipamount]

}
\posttitle{\end{center}}
\begin{document}
\maketitle

\newpage

\begin{minipage}[b]{.3\textwidth}
\includegraphics[scale=0.4]{PU_logo.eps}
\end{minipage} 
\hfill
\begin{minipage}[b]{.6\textwidth}
\begin{flushright}
\Large{\bf{Dr. Suchetana Chatterjee}}
\\ \large{Assistant Professor}
\\Deparment of Physics
\\Presidency University
\end{flushright}
\end{minipage}

\vspace{2cm}

\begin{center}
\begin{Large}
\underline {\bf {Certificate}}
\end{Large}
\end{center}

This is to certify that the work incorporated in the thesis ``Estimating the Sunyaev-Zel'dovich Signal from Quasar Hosts using a Halo Occupation Distribution based approach" has been submitted by Mr. Dhruba Dutta Chowdhury to the Department of Physics at Presidency University, Kolkata in partial fulfilment for the award of the degree of Master of Science. This report is a bonafide record of the research work carried out by Mr. Dutta Chowdhury under my supervision. To the best of my knowledge, the results presented in this report have not been submitted to any other university or institute for the award of any degree or diploma.\\\\
\begin{flushright}
Dr. Suchetana Chatterjee
\\Thesis Adviser
\end{flushright}
\today \\
Kolkata

\newpage
\vspace*{\fill}
\begin{center} 
\begin{Large}
\underline{ \bf {Declaration by the Candidate}}
\end{Large}
\end{center}
I hereby declare that the work reported in this thesis is original. It was carried out by me at the Department of Physics at Presidency University, Kolkata, India, under the supervision of Dr. Suchetana Chatterjee. The report is based on discovery of new facts and new interpretation of established facts by others. The author is solely responsible for unintentional oversights and errors, if any. I further declare that it has not formed the basis for the award of any degree, or diploma, of any other university or institution.\\\\

\begin{flushleft}
 \today
\\Department of Physics
\\Presidency University
\\86/1 College Street
\\Kolkata-700 073, West Bengal, India
\end{flushleft}
\begin{flushright}
Dhruba Dutta Chowdhury
\end{flushright}
\vspace*{\fill}

\newpage
\begin{abstract}
The Sunyaev-Zel'dovich (SZ) effect is a spectral distortion in the Cosmic Microwave Background (CMB), caused due to up-scattering of CMB photons by high energy electron distributions. The largest SZ distortion in the CMB is caused by the hot electrons present in the intra-cluster medium (ICM). However several other small scale astrophysical processes can also contribute to the SZ distortion in the CMB. 

Analytic studies have shown that the interstellar (ISM) electron gas of the host galaxy heated by quasar feedback can also cause substantial SZ effect. For successful detection of the quasar feedback signal, the SZ signal from the virialised gas in the host haloes  of quasars needs to be properly quantified. In this dissertation work, I have estimated the SZ signal from quasar hosts using analytic models of the virialised gas in the ICM/ISM. As a new extension to existing work I have used the measured Halo Occupation Distribution properties of quasar hosts.  The results show that the average SZ signal from quasar hosts decreases with redshift, varying from $10^{-4}\ {\rm arcminute}^{2}$ at $z = 0.1$ to $10^{-5}\ {\rm arcminute}^{2}$ at $z=1$. This result is consistent with what has been observed by the Planck team and will be used to theoretically validate the experimental findings of Ruan et al. (2015) who reported a first detection of the SZ signal from quasar feedback. 
\end{abstract}
\newpage
\tableofcontents
\newpage
\listoftables
\newpage
\listoffigures
\newpage
\chapter*{Acknowledgements}

I would like to take this opportunity to thank my thesis adviser Suchetana Chatterjee. It has been a great experience working with her for my master's thesis. Her constant encouragement and guidance have been instrumental in completing my dissertation. I look forward to continuing my work with her. I would also like to thank Ritaban Chatterjee, who has kindly agreed to be the co-reader of my dissertation work. Both of them have taken great pains to read through my project proposal and point out every mistake committed.

I am grateful to all the professors who have taught me, starting from my undergraduate days at Presidency College and helped to lay my foundations in physics, to my school teachers who have done all the hard work to nurture me into what I am today, and especially to my high school physics tuition teacher, Mr.\ P.K. Singh whose passionate teaching made me love physics and take it up as a career.

I would like to thank all PRESI-PACT\ (our research group) members. Our group-meetings were immensely fruitful where we learnt from each other. 

I thank all my friends who have played a major role in my life both academically and otherwise. Life without them is unimaginable. In the last two years of my formal education at Presidency University, I have made friendships that I will cherish forever. 

Last but not the least, I thank my family for having faith in me, particularly my mother who has been a great source of inspiration. Her untiring work and courage to succeed in the face of adversities encourages me to live up to her expectations and scale newer heights.

\chapter{Introduction}
In this chapter, I shall review the background on which my project is based, thereby establishing a motivation for my dissertation work. I shall also present the organisation of the rest of my thesis.

\section{Standard Model of Cosmology}
Our current understanding of the large-scale properties of the Universe is based on the standard model of cosmology, which assumes that the Universe began in an initially hot and dense state and has been expanding since then. Our paradigm relies on the cosmological principle which states that on length scales of the order of 100 Mpc or more the Universe is spatially homogeneous and isotropic. The total energy of the Universe is conserved and as the Universe expands adiabatically, its energy density and temperature decreases. Within the framework of Einstein's General Relativity and the cosmological principle, the space-time separation between two events is given by the Friedmann-Robertson-Walker (FRW) metric-
\begin{equation}
ds^{2} = -c^{2}dt^{2}+a^{2}\left(t\right)\left(dr^{2}+S^{2}_k\left(r\right)d\Omega^{2}\right)
\label{eqn1}
\end{equation}
where $a(t)$ is the dimensionless scale factor of the Universe with $a(t_0)=1$, which is the scale factor at the current epoch,
$(r,\theta,\phi)$ are the comoving coordinates of a point in space and t is the cosmological proper time. As to the contents of the Universe, it is spatially flat and about $70\%$ of the total energy density consists of dark energy, which is either the `cosmological constant' or some other form of energy with a negative pressure. The remaining $30\%$ is comprised of matter (only $4\%$ is ordinary matter, the rest being dark matter). A small fraction of the total energy is in the form of radiation: photons and neutrinos \citep{sp,plank}. To explain the growth of gravitationally collapsed structures in the Universe, it is assumed that inflation (brief period of exponential expansion) in the early Universe generated density perturbations with a nearly scale invariant spectrum involving Gaussian fluctuations.  

The Standard Model was established through years of observations and theoretical effort. In 1929, Edwin Hubble measured redshifts of nearby galaxies and found that a linear relationship exists between redshifts and the proper distances of these galaxies $z=\frac{H_0}{c}r$. `Hubble's Law' suggested an expanding Universe and the subsequent aspects of Big Bang theory. However the theory was firmly established with the discovery of the Cosmic Microwave Background (CMB) in the 1960s \citep{cmb,cmbe}. Along with the CMB and other observations, the standard model of cosmology known as the Lambda Cold Dark Matter model ($\Lambda$CDM) was established (e.g., Freedman et al. 2001; Spergel et al. 2003; Percival et al. 2007; Kowalski et al. 2008). 

\section{Cosmic Microwave Background}
The existence of the CMB was first predicted by George Gammow, Ralph Alpher and Robert Herman in 1948 based on the Big Bang Model \citep{alp}. They estimated the CMB temperature to be 5K. It was again predicted by Yakov Zel'vodich and Robert Dicke in the early 1960s and finally discovered by Arno Penzias and Robert Wilson in 1964. While working with a radio antenna in Bell Laboratories Penzias and Wilson observed an isotropic signal at $\lambda=7.35$\ cm which was free from seasonal variations and could not be associated with an isolated celestial source \citep{cmb}. In a companion letter to the same issue of the journal \citet{cmbe} proposed that the isotropic signal detected by \citet{cmb}  could be the relic radiation from the early, hot, dense and opaque universe as predicted by the Big Bang theory.

The CMB spectrum was first measured accurately by the Cosmic Background Explorer (COBE) satellite over a wide range of wavelength. COBE measured the spectrum in the wavelength range $0.1$\ mm\ $<\lambda<10$\ mm and found it to be very close to that of an ideal blackbody with an accuracy $\Delta\epsilon/\epsilon\leq10^{-4}$ at any point $(\theta,\phi)$ on the sky \citep{cobe}. The CMB temperature field has a dipolar anisotropy, which arises due to the motion of the satellite with respect to a frame in which the CMB is at rest. COBE accurately measured the dipole anisotropy and found its amplitude $\Delta{T}_{dipole}=3.372 \pm 0.004$\ mK (95\% CL) \citep{cobe1}.

After subtraction of the dipole distortion, it is found that the radiation field is nearly isotropic with mean temperature of the all sky CMB map measured to be $T_\circ =2.728\pm0.002$\ K (95\% C.L) \citep{cobe1} and relative standard deviation of 1 part in $10^{5}$ \citep{cobe}. COBE had an angular resolution of $7^\circ$.

\begin{equation}
\left<T\right>=\frac{1}{4\pi}\int T\left(\theta,\phi\right) \sin \theta d\theta d\phi=2.728K
\label{eqn2}
\end{equation}
\begin{equation}
\left<\left(\frac{\delta{T}}{T}\right)^{2}\right>^{1/2}=1.1\times 10^{-5}
\label{eqn3}
\end{equation}

With a better resolution of $0.3^\circ$, the Wilkinson Microwave Anisotropy Probe (WMAP) has measured some of the basic parameters of the $\Lambda$CDM model with very high precision \citep[e.g.][]{sp, du}. These parameters are  dark matter density in the Universe ($\Omega_{DM}$), baryonic matter density ($\Omega_{b}$), the Hubble constant ($H_0$), the scale dependence of fluctuations ($n_{s}$), and the redshift of reionization ($z_{reion}$). These together with other measurements, determine the remaining parameters of the standard model.

Coming to the state of the art measurements, the {\it Planck} satellite has a resolution of $5-33'$ (depending on the frequency channel). The first {\it Planck} data was released in 2013 and it further constrained the cosmological parameters \citep{plank}. According to its 2013 data release, the universe is $13.798 \pm 0.037$ billion years old, and contains $69 \pm 1\%$ dark energy, $25.8 \pm 0.4 \%$ dark matter and $4.82 \pm 0.05 \%$ ordinary matter. The Hubble constant is found to be $67.4 \pm 1.4\ {\rm(km/s)/Mpc}.$ \citep{plank}.

\begin{table}[h!!!!!!]
\centering
\begin{tabular}{|c|c|}
\hline
Cosmological parameters & Best fit value (68 \% C.L.)\\
\hline
Dark Enegry density, $\Omega_{\Lambda}$ & 0.67 $\pm$ 0.02 \\
Dark matter density, $\Omega_{dm}$ & 0.27 $\pm$ 0.02 \\
Baryon density, $\Omega_{b}$ & 0.04 $\pm$ 0.01\\
Matter density, $\Omega_{m}$ & 0.31 $\pm$ 0.02\\
Hubble constant, H & 67.4 $\pm$ 1.4 km/s/Mpc \\
Reduced Hubble constant, h=H/100 & 0.67 $\pm$ 0.01 \\ \hline
\end{tabular}
\caption[Cosmological Parameters]{Cosmological Parameters as determined by \citet{plank}}
\end{table}

\newpage
\begin{figure}[h!!!]
\centerline{
\epsfig{figure=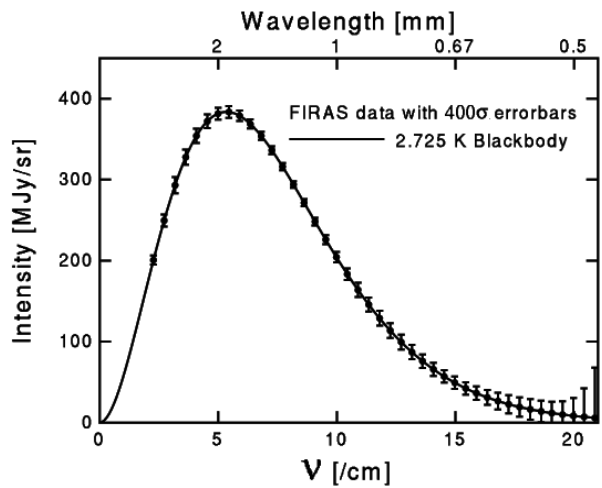,width=8.0cm,angle=0.}}
\caption[COBE blackbody spectrum]{Blackbody spectrum from COBE at 2.728 K. Image courtesy: COBE science team \citep{cobe1}
}
\label{fig1}
\end{figure}
\begin{figure}[h!!!!!!!!!!!!!!!!!!!!]
\centerline{
\epsfig{figure=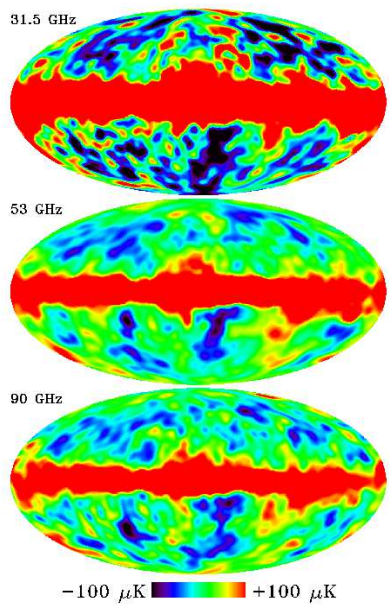,width=7.0cm,angle=0.}}
\caption[COBE all-sky map]{Four-year temperature smoothed all sky maps from COBE over angular resolution of $10^\circ$ after subtraction of the monopole and dipole anisotropy. Image courtesy: COBE science team \citep{nasa1}
} 
\label{fig2}
\end{figure}

\begin{figure}[h!!!!!!!!!!!]
\centerline{
\epsfig{figure=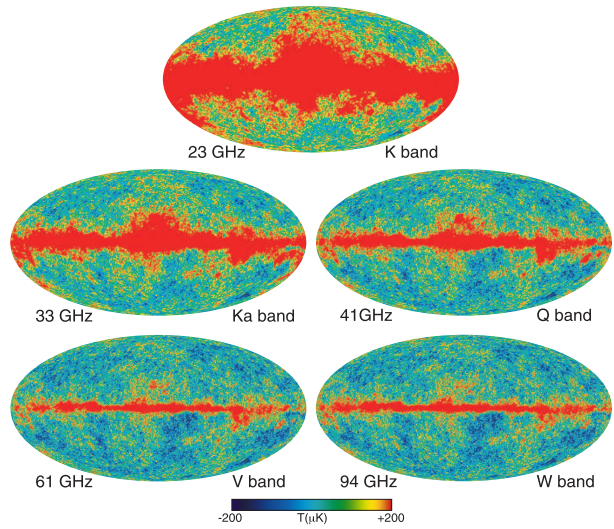,width=8.0cm,angle=0.}}
\caption[WMAP all-sky map]{WMAP Nine-year temperature all sky maps measured to an angular resolution of $0.3^\circ$ shown in a Mollweide projection after subtraction of the monopole and dipole anisotropy. The anisotropies are resolved further compared to COBE. Image courtesy: WMAP science team \citep{wmap}
}
\label{fig3}
\end{figure}
\newpage

\begin{figure}[h!!!!!!!!!!!]
\centerline{
\scalebox{0.50}{\includegraphics{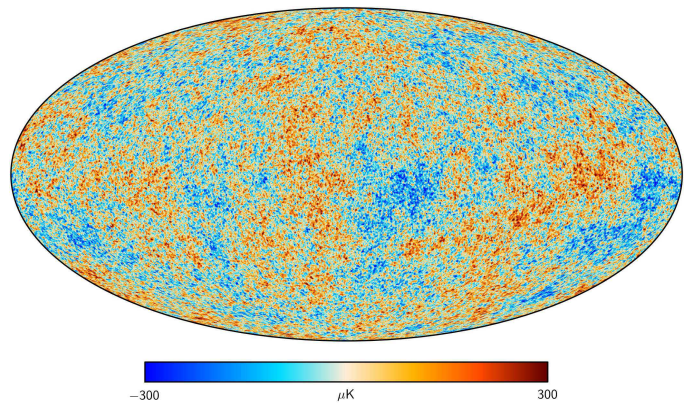}}}
\caption[{\it Planck} all-sky map]{Planck SMICA February (2015) all sky map of temperature anisotropies after subtraction of the monopole, dipole and galactic dust correction. Planck has a resolution of $5-33'$ depending on the frequency channel. Image courtesy: ESA, Planck Collaboration 
}
\label{fig4}
\end{figure}
\newpage
\begin{figure}[h!!!!!!!!!!]
\centerline{
\epsfig{figure=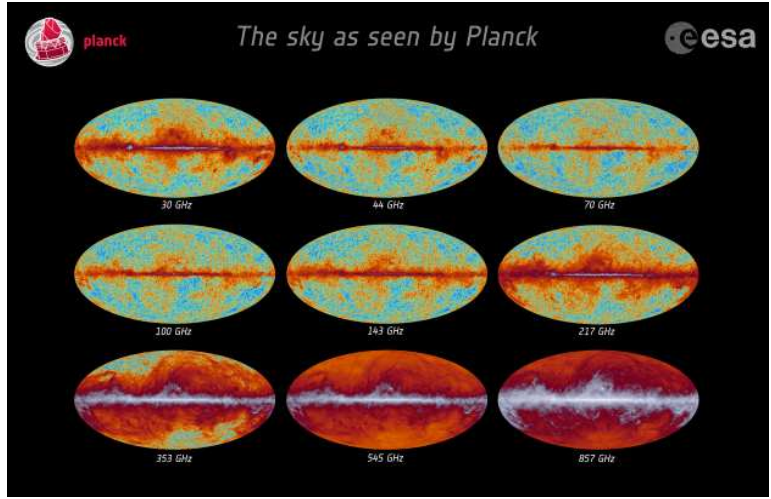,width=10.0cm,angle=0.}}
\caption[{\it Planck} all-sky map at all its nine frequencies]{Planck all-sky maps at nine frequencies during its first 15.5 months of observations released in March 2013. The Cosmic Microwave Background is most evident in the frequency bands between 70 and 217 GHz. Observations at the lowest frequencies are affected by foreground radio emission from the interstellar material in the Milky Way, mostly due to synchrotron radiation emitted by electrons. At the highest frequencies, observations are affected by emission from interstellar dust in the Milky Way. The combination of data collected at all the nine frequencies is important for optimal reconstruction of the noise, in order to subtract them and find the exact signal from the CMB. Image courtesy: ESA, Planck Collaboration 
}
\label{fig5}
\end{figure}

\subsection{Origin of the CMB}

In the hot and dense initial stages of the Universe, the average energy per photon was very high and with a very large photon to baryon ratio, any atom which formed by radiative combination of protons and electrons was short-lived. Thus the Universe was almost completely ionised. Due to frequent Thomson scatterings of the photons with the electrons and proton-electron interaction, the baryon-photon plasma was tightly coupled, rendering the Universe opaque. Any density fluctuations in the plasma at scales smaller than the horizon size tended to smoothen out, resulting in a homogeneous Universe (at least at scales of the horizon size) in thermal equilibrium, with the energy density of the photons being nearly that of a blackbody.

Then as the Universe expanded adiabatically, the temperature of the plasma fell. The ionization fraction of the Universe which is a measure of the number density of free electrons started decreasing and stable atoms began to form. At about T\ =\ 3740K,the ionization fraction reduced to half, there were not enough partners left for the photons to scatter from and the plasma began to go out of equilibrium. Finally when the expansion rate of the Universe overtook the Thomson scattering rate, the photons decoupled from the plasma and the Universe became transparent. Most of the photons could no longer scatter off the remaining free electrons and the epoch of last scattering followed soon after. This is the epoch in the history of the Universe when a typical photon underwent scattering for the last time and has been free-streaming since then. In the $\Lambda$CDM model, ( $\Omega^\circ_\Lambda=0.7, \Omega^\circ_m=0.3, \Omega^\circ_b=0.04, \Omega^\circ=1, \Omega^\circ_r=8.5\times10^{-5}$ ) and the epoch of last scattering occurs at z\ $=1100$, when the temperature of the Universe is $3000$\ K. It is these photons that comprise the Cosmic Microwave Background. Since then the Universe has expanded and CMB photons have been continuously redshifted to lower energies, maintaining a blackbody spectrum (excluding late Universe effects on the CMB) following the relation
\begin{equation}
T(z)=T_\circ(1+z)
\label{eqn4}
\end{equation}
Following Wein’s displacement law the peak of the CMB spectrum occurs around $\lambda_{max}\approx 2\ {\rm mm}$, $\nu_{max}\approx {\rm160}\ GHz$ with mean energy density $\epsilon^\circ_r=\alpha T_\circ^{4}=0.261\ {\rm MeV m^{3}}$. The near isotropic blackbody spectrum of the CMB as observed at the current epoch provides strong evidence in favour of the Big Bang Model. 

It has been already noted that the CMB temperature field is not perfectly isotropic. The anisotropies of the CMB can be broadly divided into two categories namely primary anisotropies due to effects which occurred at the epoch of last scattering or before and secondary anisotropies due to late time effects on the CMB photons during their journey from the last scattering surface to the observer.

Among the secondary anisotropies important ones are the gravitational lensing of the photons by late Universe dark matter haloes and Sunyaev Zel'dovich Effect which occurs due to scattering of the CMB photons by rich concentrations of hot electrons in the Universe (e.g., galaxy clusters). While the primary anisotropies tell us about the seeds of structure formation in the Universe, the secondary anisotropies throw light on the properties of the structures in the late Universe. Together they can be used as a probe of structure formation in the Universe.

\section{Sunyaev-Zel'dovich Effect}

The most prominent of the secondary anisotropies of the CMB is the Sunyaev-Zel'dovich (SZ) effect \citep{sz,sz1}. It is further classified into Thermal and Kinetic SZ effect.

\subsection{Thermal Sunyaev-Zel'dovich Effect}

Thermal Sunyaev-Zel'dovich (TSZ) effect is a spectral distortion of the CMB due to inverse Compton scattering of CMB photons by a distribution of hot free electrons with $T_{e} >> T_{CMB}$, where $T_{e}$ is the temperature of the elecron distribution. The effect was first predicted by \citet{sz}. The theoretical derivation of the TSZ temperature fluctuation assuming a non-relativistic electron distribution can be found in the literature \citep{sz,sz1,re}. Relativistic corrections to it can be found in e.g., Rephaeli (1995).

In this thesis I have followed the derivation from \citet{thesis} and \citet{book1} where we start from the collisional Boltzman equation. Since the electrons are non-relativistic ($K_bT_e<<m_ec^{2}$), the scattering cross-section can be approximated by Thompson scattering. The main features of the theoretical calculation are presented below. 

\begin{enumerate}
\item On an average the photons gain energy on scattering and move from low to high $\nu$. As photons are neither created nor destroyed in scattering events, we observe a fall in intensity below and a rise of intensity above a certain critical frequency, at which the intensity of the SZ spectrum is exactly similar to the CMB blackbody at a given temperature. It is called the null frequency and comes out to be $\sim 218$\ GHz \citep{c}. This is an unique observational signature of the TSZ effect.\\\newline

\item The change in intensity from the CMB blackbody, $\Delta{I}_{SZ}$ is given by-
\begin{equation}
\Delta{I}_{SZ}=g(x)I_0y
\end{equation}
 where $I_0=2(k_bT_{CMB})^{3}/(hc)^{2}$,\ $x=h\nu/k_bT_{CMB}$ and
\begin{equation}
g(x)=\frac{x^{4}e^{x}}{(e^{x}-1)^{2}}\left(x\frac{e^{x}+1}{e^{x}-1}-4\right)
\end{equation}
\begin{equation}
y_l=\frac{\sigma_T}{m_e c^{2}}\int n_e k_b T_e dl
\label{y}
\end{equation}
$\sigma_T$ is the Thompson scattering cross-section, $m_ec^{2}$ is the electron rest-mass energy, $n_e$ is the electron number density, $k_b$ is Boltzman constant and $T_e$ is the temperature of the electron. $y_l$ is called the Compton $y$ parameter and is proportional to the line of sight integral of the thermal pressure $P_e=n_ek_bT_e$ of the electron distribution.\\\newline

\item If one associates a brightness temperature with the change of intensity, an increase and decrease of intensity manifest as hot and cold spots in the CMB map.
\begin{equation}
\Delta{T}_{SZ}/T_{CMB}=f(x)y
\label{sz}
\end{equation}
where 
\begin{equation}
f(x)=\left(x\frac{e^{x}+1}{e^{x}-1}-4\right)
\end{equation}
\end{enumerate}

One of the richest sources of high energy electron distribution in the Universe is the intra-cluster medium (ICM) of galaxy clusters. The ICM consists of virialised electron gas, having typical temperatures of the order of $10^{7}-10^{8}$\ K. A quick calculation shows that $K_bT_e\ \sim\ 10^{-2}$ Mev much less than the electron rest mass energy of 0.51 Mev. Thus it can be safely assumed to be non-relativistic and the above formalism can be applied. The integrated SZ effect from a cluster $Y_{in}$ is proportional to \citep{c}.
\begin{equation}
\int y_ld\Omega   \ \alpha  \ \int n_e k_b T_e dl d\Omega   \ \alpha  \ N_e<T>/D_A^{2}  \ \alpha  \ M_g<T>/D_A^{2}
\end{equation}
where $d\Omega$ is the solid angle subtended by the cluster, $D_A$ is its angular diameter distance, $N_e$ is the total electron number of the distribution, $<T>$ is the average temperature of the cluster and $M_g$ is the mass of the cluster gas. The angular diameter distance $D_A$ does not vary much at high redshift, and a cluster of a given mass is denser and hence hotter at high redshift as the mean matter density increases as $(1 + z)^{3}$. So the SZ signal has very little dependence on redshift and clusters above a certain mass limit can be detected with equal ease at low and high redshits \citep{c}. However since the ICM is diffuse, $Y_{in}$ for typical clusters is of the order of $\sim 10^{-4}$, producing $\Delta{T}\sim$ 1 mK \citep{c}. It is thus an important tool in cluster cosmology and can be used to study evolution of galaxy clusters as a function of redshift, providing hints about structure formation and evolution.

\subsection{Kinetic Sunyaev-Zeldovich Effect}

A spectral distortion of the CMB also arises if the electron distribution (source of scattering) has a line of sight bulk velocity, and the CMB spectrum suffers a net shift due to the Doppler effect \citep{c}. This is called the Kinetic Sunyaev-Zel'dovich effect. It can be used to measure peculiar cluster velocities. Thermal SZ effect is in general more dominant of the two.
\vspace{2cm}

\begin{figure}[h!!!!]
\centerline{
\epsfig{figure=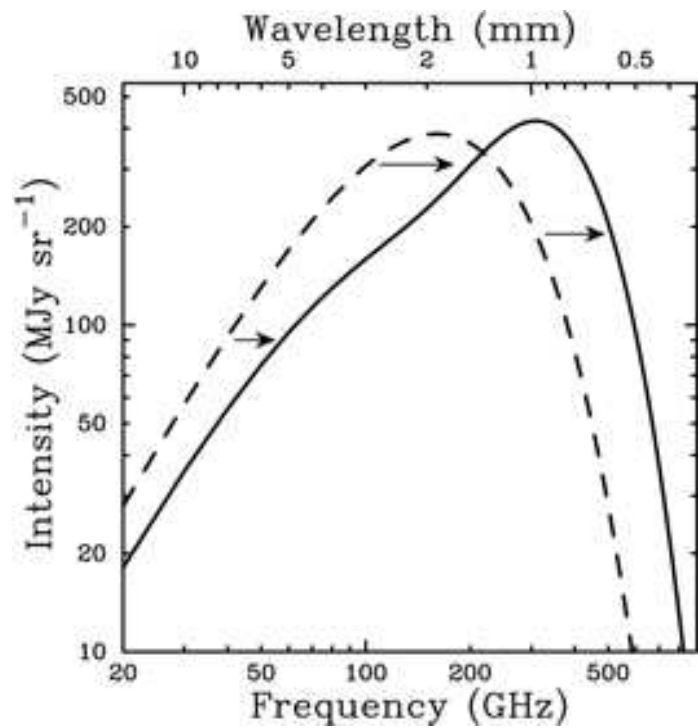,width=9.0cm,angle=0.}}
\caption[Spectral distortion from SZ effect]{The dashed line shows blackbody spectrum of CMB Photons before scattering which is modified after Sunyaev Zel'dovich Effect from a hot cluster.The modified spectrum is shown by the solid line due to a fictional cluster 1000 times more massive than original galaxy clusters.The existence of null frequency at $\sim 218$ GHz is evident. Image credit: \citet{c}
}
\label{fig6}
\end{figure}

\begin{figure}[h!!!!]
\centerline{
\epsfig{figure=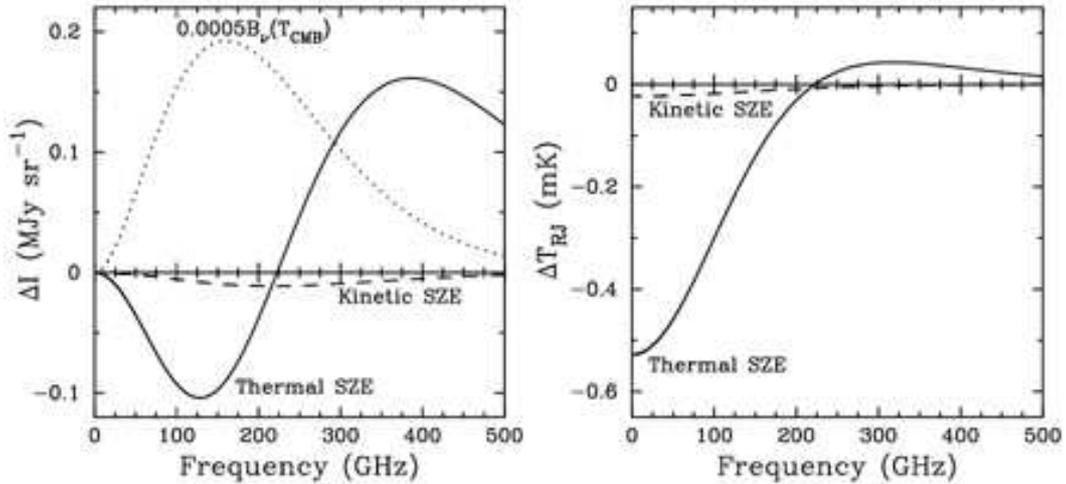,width=14.0cm,angle=0.}}
\caption[Comparison of Thermal and Kinetic SZ effect]{The left panel shows the variation of $\Delta I_{SZ}$ and the right panel  variation of $\Delta T_{SZ}$ as a function of frequency for a cluster of average temperature 10 KeV,Compton parameter $y_{in}=10^{-4}$ and peculiar velocity 500 km/sec.The CMB blackbody at 2.7K with intensity scaled down by 0.0005 is shown for reference by dotted line.The solid curve represents Thermal SZ effect and the dashed curve is due to Kinetc SZ effect. Image credit: \citet{c}
}
\label{fig7}
\end{figure}

\newpage

\section{Quasars and Thermal Sunyaev-Zel'dovich Effect}

In this section, I will give a brief description of quasars and feedback processes in a galaxy, thereby trying to establish the motivation for my dissertation.

It is believed that all massive galaxies have a supper massive black-hole (SMBH) at their centres with masses in the range $10^{6}-10^{9}M_\odot$ \citep[e.g.,][]{oo,sol,tre}. About $1-10\%$ of these are active \citep[e.g.][]{sol,dan}, meaning they are accreting matter from the surroundings at a rate strong enough to produce high luminosity for the central region of the galaxy. Such galaxies are said to possess an Active Galactic Nuclei (AGN) and are referred as AGN host galaxies.

\subsection{Quasar Feedback}

The most luminous AGNs are called quasars with luminosities of the order of $10^{45}-10^{47} erg s^{-1}$, which is close to the Eddington limit for the SMBHs. The Eddington Limit is the maximum luminosity of an AGN at which the outward radiation pressure on the accreting matter balances the inward gravitational pull of the black hole. The Sloan Digital Sky Survey (SDSS) has detected over one million quasars through its photometric survey \citep{ric1} and 308,377 quasars have optical spectra till its Data Release 10 with their comoving space density peaking at $z\sim2$) \citep{boss,sdss}. An interesting observation is that quasars were more common in the past and are very rare in nearby galaxies. Apparently most SMBHs in the present day are accreting at low or moderate rates, and hence are producing less luminosities than the Eddington luminosity $L_E$. The rise and fall of the quasar population, in addition to the relation between SMBH masses and the velocity dispersion of host galaxy bulges ($M_{BH}-\sigma$ relation) suggest a likely connection between the growth and evolution of galaxies with their central SMBH \citep{f,me,g}.

Accretion in Quasars is caused due to large scale inflows of gas  towards the black hole, possibly due to galaxy mergers \citep{b,s,h}. While the size of the accretion disc is of the order of few parsecs, as the accretion continues energetic outflows in the form of winds or as relativistic jets are launched from the accretion disk surrounding the black hole. This is called quaser feedback which injects significant amount of kinetic energy into the interstellar medium (ISM) (Kpc scale) and the surrounding circumgalactic region, resulting in removal of gas from the host galaxy and thus limiting the gas inflow into the SMBH regulating its accretion rate \citep{fab,sil}. This can also terminate star formation in the host galaxy bulge by stopping gas cooling \citep[e.g.,][]{far,pa,la}. 

Hence quasar feedback may hold the key to correlations between SMBH mass and host galaxy properties, redshift evolution of star formation and quasar activity in galaxies. While many of these observed trends have been reproduced in semi-analytical models and cosmological simulations of galaxy formation \citep[e.g.][]{ka,di,bo,spe,si,so}, details of the feedback energetics, mechanism, and effects on galaxy evolution are still largely unknown, as quasar activity mainly occurs at high redshifts and is often obscured by dust and gas \citep[e.g.,][]{mao,oh,gra,ruan1}.

\subsection{Thermal Sunyaev-Zel'dovich Effect as a Probe of Feedback}

Since quasar feedback alters the thermal energy of the free electrons in the ISM of the host galaxy and the surrounding circumgalactic medium, thermal Sunyaev Zel'dovich effect can be used as a potential probe to understand this feedback process. As TSZ signal is weakly dependent on redshift, it can be used to detect quasar feedback upto very high redshifts. However the feedback signal predicted by anatytic studies is too low to be detected for individual quasars \citep[e.g.,][]{sc07,sc08,lp,na}. An additional complication is that the TSZ signal measured will be in general coming from both the ionized gas in the host galaxy heated from quasar feedback and the ICM of the host halo in which the quasar resides. The signal from the host halo ICM can also get modified due to input of feedback energy. Thus a proper calibration of the SZ signal from the ICM in absence of any feedback mechanism is necessary so that if a significant alteration in it within the limits of experimental accuracy is obtained, a detection of SZ signal due to quasar feedback can be claimed.

\section{Chapter Organisation} 
In this dissertation work, I have estimated the SZ signal from the ICM of quasar host haloes, using analytic models of the virialised gas in the ICM and through measured Halo Occupation Distribution (HOD) properties of quasar hosts. In Chapter 2, I shall present a model for the halo ICM which can predict average properties of the halo gas. This model shall be used in Chapter 3 to calculate the total SZ distortion from a dark matter halo. Its dependence on halo mass and redshift will be studied. Thereafter in Chapter 4, I shall estimate the average SZ signal from all quasar hosting haloes at a given redshift using the Halo Mass Function and the HOD to model the number of quasar hosting haloes at a given redshift. Since individual halo signals are weak, the average signal has to be studied and only a statistical detection of the signature of quasar feedback in the SZ signal can be claimed if any. Finally in Chapter 5, I shall elucidate on the future goals of this project. Very recently Ruan et al.\ (2015) have claimed to have detected the signature of quasar feedback in the SZ signal using Planck data. As an extension of my work, I shall compare my theoretical model with the findings of Ruan et al.\ (2015). 

In my work I have used a $\Lambda$CDM Cosmology with cosmological parameters taken from Planck Collaboration et al. XVI (2014). The values of the relevant parameters are given in Table 1.1.

\chapter{Modelling the Intracluster Gas in Quasar Hosting Haloes}

In this chapter, I will discuss a model for the virialised gas in a dark matter halo following the prescription of Komatsu \& Seljak (2001), (2002). The model predicts the radial profiles of the density and temperature of gas inside a virialised halo under simple assumptions motivated from numerical simulations. The density and temperature will be used in chapter 3 to calculate the Compton y-parameter from the halo.

\section{Universal Dark Matter Density Profile}
As suggested by high-resolution N-body simulations (Navarro et al. 1996, 1997; Moore et al. 1998) the dark matter density profile in a halo is well described by a self similar spherically symmetric form. This means that the dark matter density profile can be represented by a dimensionless function, $y_{dm}(r)$ so that
\begin{align}
\rho_{dm}\left(r\right) &= \rho_{s} \ y_{dm}\left(r/r_s\right) \\
x &= r/r_s
\end{align}
where $\rho_s$ is the mass density at a characteristic radius $r_s$ from the centre of the halo which serves as a normalisation factor.

The dark matter mass within a sphere of radius r is given by -
\begin{align}
M\left(\leq r\right) &= \int_{0}^{r} \rho_{dm}\left(r\right) 4\pi r^2 dr \\
 &= 4\pi \rho_s r_s^3 \int_{0}^{x} y_{dm}(u)u^2 du, \hspace{0.5cm} u=r/r_{s} \nonumber \\
 &= 4 \pi \rho_s r_s^3 m(x) \\
 m(x) &=\int_{0}^{x} y_{dm}(u)u^2 du
\end{align}

In the current work the total dark matter mass of the halo, $M$ is defined as the mass within the virial radius, $r_{vir}$ of the halo, $M \equiv M (\leq r_{	vir})$.
$c=r_{vir}/r_s$ is called the concentration parameter of the halo. From Eq.\ 2.4, for $x=c$ we get
\begin{equation}
\rho_s = \frac{c^3 M}{4 \pi r_{vir}^3 m(c)}
\end{equation}

The virial radius of a halo is calculated according to the spherical collapse model (Press \& Schechter 1974; Peebles 1980).
\begin{equation}
r_{vir}=\left(\frac{M}{4/3\pi \Delta_c(z) \rho_c(z)}\right)^{1/3}
\end{equation}
where $\Delta_c(z)$ is the spherical overdensity of the virialised halo within $r_{vir}$ at redshift z in units of the critical density of the universe, $\rho_c(z)$ given by (e.g., Bryan \& Norman 1998)
\begin{align}
\Delta_c(z) &= 18 \pi^2 + 82 y - 39 y^2 \\
y &= \frac{\Omega_m^0 (1+z)^3}{\Omega_m^0 (1+z)^3 + \Omega_{\Lambda}}-1 \hspace{0.5cm} for \ \Omega_r= 0
\end{align}
where $\Omega_m^0$ is matter density of the universe at z=0 and $\Omega_{\Lambda}$ is the dark energy density (cosmological constant) as a fraction of the critical density at $z=0$. 

The concentration parameter c is given by 
\begin{align}
c &= \frac{c_0}{1+z} \\
c_0 &= 6\left(\frac{M}{10^{14} M_\odot}\right)^{-1/5}
\end{align}
where $c_0$ is taken from Seljak (2000) and the redshift dependence of $c$ is taken from Bullock et al.\ (2001). Mass of the halo is in solar mass units.

The functional form of $y_{dm}(x)$ is taken to be 
\begin{equation}
y_{dm}(x)=\frac{1}{x^{\alpha}(1+x)^{3-\alpha}}
\end{equation}
For $x \gg 1$, $y_{dm}(x) = x^{-3}$ which is the average behaviour found in most simulations (eg. Navarro et al 1996,1997; Moore et al. 1998; Thomas et al. 2001).
For $x \ll 1$, $y_{dm}(x) = x^{-\alpha}$ where $1 < \alpha < 3/2$ (Navarro et al. 1996, 1997; Moore et al. 1998). I will use $\alpha = 1$ in my work which corresponds to the Navarro, Frenk \& White (NFW) profile proposed by Navarro et al. (1996) (1997). Thus 
\begin{equation}
\rho_{dm}(x)= \frac{\rho_s}{x(1+x)^{2}}
\end{equation}

For NFW profile,
\begin{align}
m(x) = \int_{0}^{x} y_{dm}(u)u^2 du &= {\rm ln} (1+x) - \frac{x}{(1+x)} \\
\int_{0}^{x} \frac{m(u)}{u^2} du &= 1 - \frac{{\rm ln}(1+x)}{x}
\end{align}

\section{Universal Halo Gas Profile }

\subsection{Polytropic Gas Model in Hydrostatic Equilibrium}
If no additional scale-dependence is introduced, the gas density profile in a halo would also be self similar following the assumption for the dark matter density profile.
\begin{equation}
\rho_{gas}(r)=\rho_{gas}(0)y_{gas}(r/r_s)
\end{equation}
whrere $\rho_{gas}(0)$ is the gas density at $r=0$.

To take into account the effect of gas temperature gradient which is found from both observations and simulations a polytropic equation of state is assumed following Suto et al. (1998).
\begin{equation}
P_{gas}(r) \hspace{0.2cm} \alpha \hspace{0.2cm} \rho_{gas}(r)T_{gas}(r) \hspace{0.2cm} \alpha \hspace{0.2cm} \rho^\gamma_{gas}(r),
\end{equation}
where $\gamma$ is the polytropic index of the gas. This gives a self similar temperature profile as follows-
\begin{align}
T_{gas} \hspace{0.2cm} &\alpha \hspace{0.2cm} \rho^{\gamma-1}_{gas} \\
T_{gas}(r) &= T_{gas}(0) \hspace{0.1cm} y^{\gamma-1}_{gas}(r/r_s)
\end{align}
where $T_{gas}(0)$ is the central gas temperature.

The gas is assumed to be in hydrostatic equilibrium, with the gas pressure gradient balancing the gravitational attraction of the dark matter. Thus
\begin{equation}
\frac{dP_{gas}}{dr} = -\frac{G\rho_{gas}M(\leq r)}{r^2}
\end{equation}
Substituting $P_{gas}$ from Eq.\ 2.17, we get the differential equation for the gas profile $y_{gas}(r/r_s)$ as
\begin{equation}
\frac{dy^{\gamma-1}_{gas}}{dr} = -\left(\frac{\gamma-1}{\gamma}\right) \frac{G \mu m_p M}{k_b T_{gas}(0)r^{2}}\frac{m(r/r_s)}{m(c)}
\end{equation}
Here $G$ is the universal gravitational constant, $\mu$ is the average mass of the gas particles as a fraction of the proton mass, and $k_b$ is the Boltzman constant. Since $y_{gas}(0)=1$ (Eq.\ 2.16), Eq.\ 2.21 can be analytically solved to give 
\begin{align}
y^{\gamma-1}_{gas}(x) &= 1- 3 \eta^{-1}(0)\left(\frac{\gamma-1}{\gamma}\right)\left[\frac{c}{m(c)}\right]\int_0^x \frac{m(u)}{u^2}du \\
\eta^{-1}(x) &= \frac{G\mu m_pM}{3 r_{vir} k_b T_{gas}(x)}
\end{align} 

\subsection{Gas tracing Dark Matter in Halo Outskirts}
To determine $y_{gas}(x)$ we require to fix two free parameters in the model namely $\eta(0)$ called the mass-temperature normalisation factor and the polytropic index $\gamma$. This is done by requiring that the gas traces the dark matter in the outer parts of the halo for $r > r_{vir}/2$ which is found in many hydrodynamic simulations (e.g., Navarro et al. 1995; Bryan \& Norman 1998). Mathematically,
\begin{align}
y_{gas}(x) \hspace{0.2cm} &\alpha \hspace{0.2cm} y_{dm}(x) \hspace{1.6cm} for \ x > c/2 \\
{\rm or}, \ \frac{{\rm d \ ln} \ y_{dm}(x)}{{\rm d\ ln } \ x} &= \frac{{\rm d \ ln} \ y_{gas}(x)}{{\rm d \ ln} \ x} \hspace{0.5cm} for\  x > c/2
\end{align}
Imposing the above condition gives the best-fit values for $\gamma$ and $\eta(0)$ as 
\begin{align}
\gamma &= 1.15+ 0.01(c-6.5) \\
\eta(0) &= 0.00676(c-6.5)^2+ 0.206(c-6.5)+2.48
\end{align}
valid for $1<c<25$ (Komatsu \& Seljak 2001).
Since $c$ depends on the mass of the halo $M$ and its redshift $z$ (Eq.\ 2.10), $\gamma$ and $\eta(0)$ are functions of halo mass and redshift.

\subsection{Central Gas Density and Temperature}
As the halo merges with the cosmic background at around the virial radius, the ratio of the matter and dark matter density should be the same as that of the cosmic average at $r=r_{vir}$. Thus the gas density is normalised as
\begin{equation}
\rho_{gas}(c)=\rho_{gas}(0)y_{gas}(c)= \frac{\Omega_{b}}{\Omega_{dm}} \rho_{dm}(c)
\end{equation}
Evaluating Eq.\ 2.13 at $x=c$ and substituting $\rho_s$ from Eq.\ 2.6 we get
\begin{equation}
\rho_{gas}(0)=\frac{1}{y_{gas}(c)} \frac{\Omega_{b}}{\Omega_{dm}} \frac{M}{4\pi r^3_{vir}} \frac{c^2}{(1+c)^2}\left[{\rm ln}(1+c)-\frac{c}{1+c}\right]^{-1}
\end{equation}

Let the density of hydrogen be $\rho_{H}$ and that of helium be $\rho_{He}$. We define their mass abundances as $X \equiv \rho_{H}/\rho_{gas}$ and $Y \equiv \rho_{He}/\rho_{gas}$. Since the halo is assumed to consist of only hydrogen and helium, $Y= 1-X$. If the number density of hydrogen atoms is $n_{H}$ and that of helium atoms is $n_{He}$, then $n_{H}=\frac {X\rho_{gas}}{m_{H}}$ and $n_{He}=\frac{Y\rho_{gas}}{4m_{H}}$. As the gases are ionised, the total number of gas particles is given by 

\begin{align}
n &= 2n_{H}+3n_{He} \nonumber \\
&= \frac{\rho_{gas}}{m_{H}}\left(2X+\frac{3}{4}(1-X)\right) \nonumber \\
&= \frac{\rho_{gas}\left(5X+3\right)}{4m_{H}} \nonumber \\
\end{align} 

Thus the average particle mass as a fraction of proton mass is $\mu = \frac{\rho_{gas}}{n\ m_{H}}=\frac{4}{5X+3}$, where we take $X= 0.75$. Hence from Eq.\ 2.23

\begin{equation}
T_{gas}(0)= \eta(0) \frac{4}{3+5X}\frac{Gm_pM}{3r_{vir}k_b}
\end{equation}

\subsection{Gas Pressure Profile}
From the ideal gas equation-
\begin{align}
P_{gas}(r) &= \frac{\rho_{gas}(r)}{\mu m_p} \ k_b T_{gas}(r) \\
&= \frac{5X+3}{4 m_p}\ \rho_{gas}(r)\ k_b T_{gas}(r)  \\
&=\frac{5X+3}{4 m_p}\ \rho_{gas}(0)\ y_{gas}(r/r_s)\ k_b T_{gas}(0)\ y^{\gamma-1}_{gas}(r/r_s) \nonumber \\
&=\frac{5X+3}{4 m_p}\ \rho_{gas}(0)\ k_b T_{gas}(0)\ y^{\gamma}_{gas}(r/r_s)
\end{align}

Thus starting from a universal dark matter profile (NFW profile here), a universal gas profile has been anatytically obtained. It models the gas density, temperature and pressure in a halo as a function of halo mass and redshift without containing any other free parameters, thereby representing the average behaviour of gas in a halo. The plots for the density, pressure and temperature profiles are shown in Figures 2.1, 2.2, 2.3 and 2.4.

\newpage
\begin{figure}[h!!!!!]

\begin{subfigure}[c]{0.5\textwidth}
\centering
\scalebox{0.92}{\includegraphics[trim={0 7cm 0 0},clip]{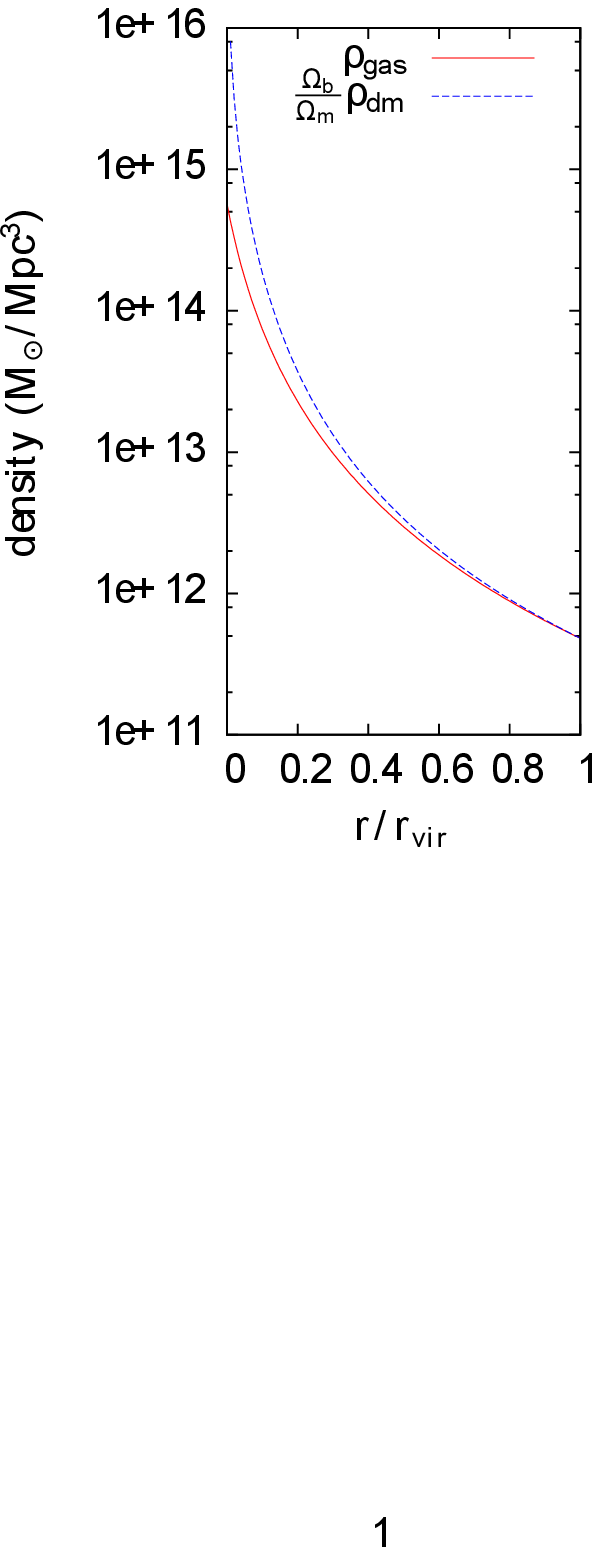}}
\end{subfigure}
\begin{subfigure}[c]{0.5\textwidth}
\centering
\scalebox{0.92}{\includegraphics[trim={0 7cm 0 0},clip]{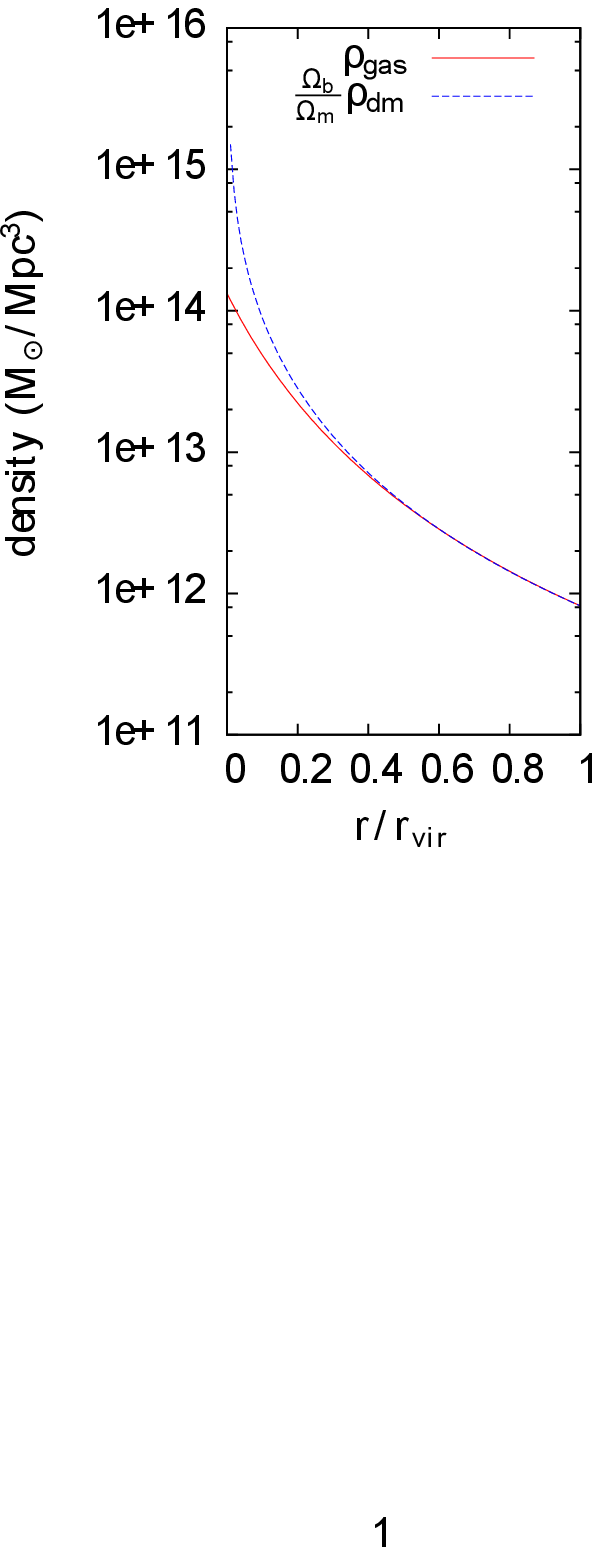}}
\end{subfigure}
\caption[Comparison of gas density and dark matter density profiles]{The gas density and dark matter density profiles for a halo of mass $10^{12}M_\odot$(left image) and $10^{15}M_\odot$(right image) at z= 0.1. The gas traces the dark matter in the halo outskrits}
\end{figure}

\begin{figure}[h!!!]
\begin{subfigure}[c]{.5\textwidth}
\centering
  \scalebox{0.92}{\includegraphics[trim={0 7cm 0 0},clip]{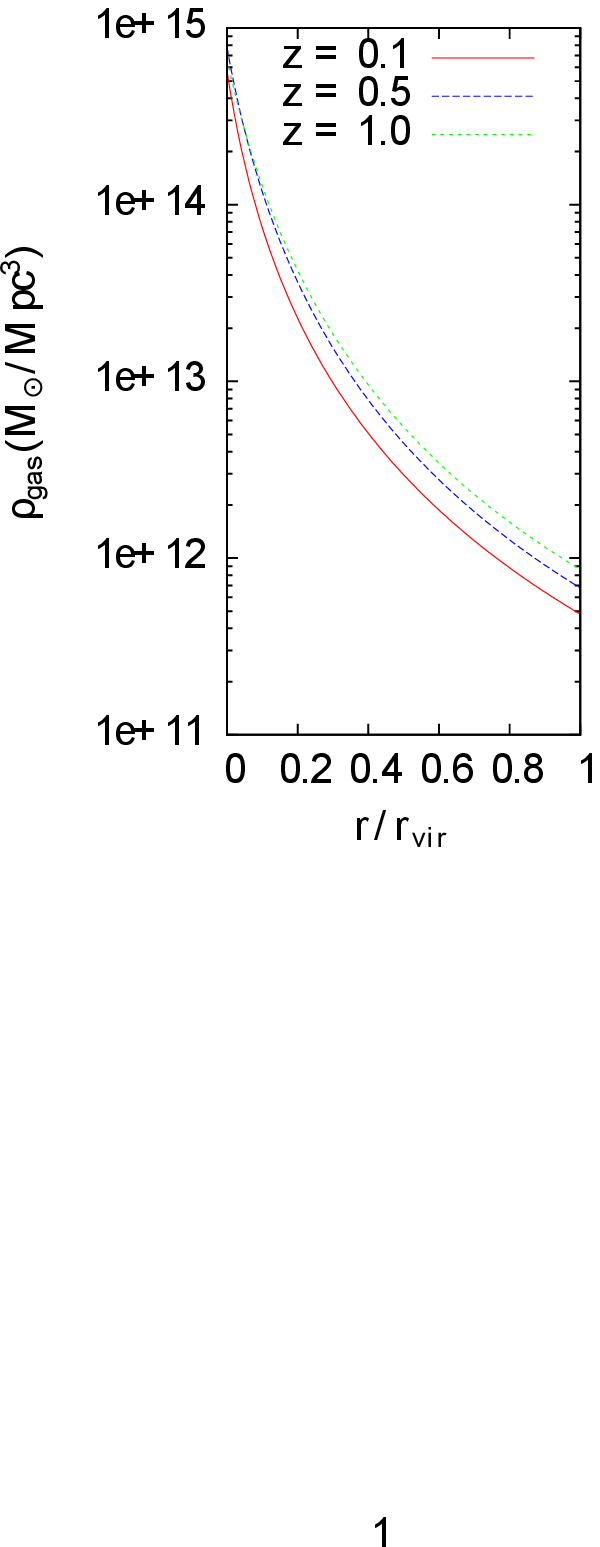}}
  \end{subfigure}
\begin{subfigure}[c]{.5\textwidth}
\centering
 \scalebox{0.92}{\includegraphics[trim={0 7cm 0 0}]{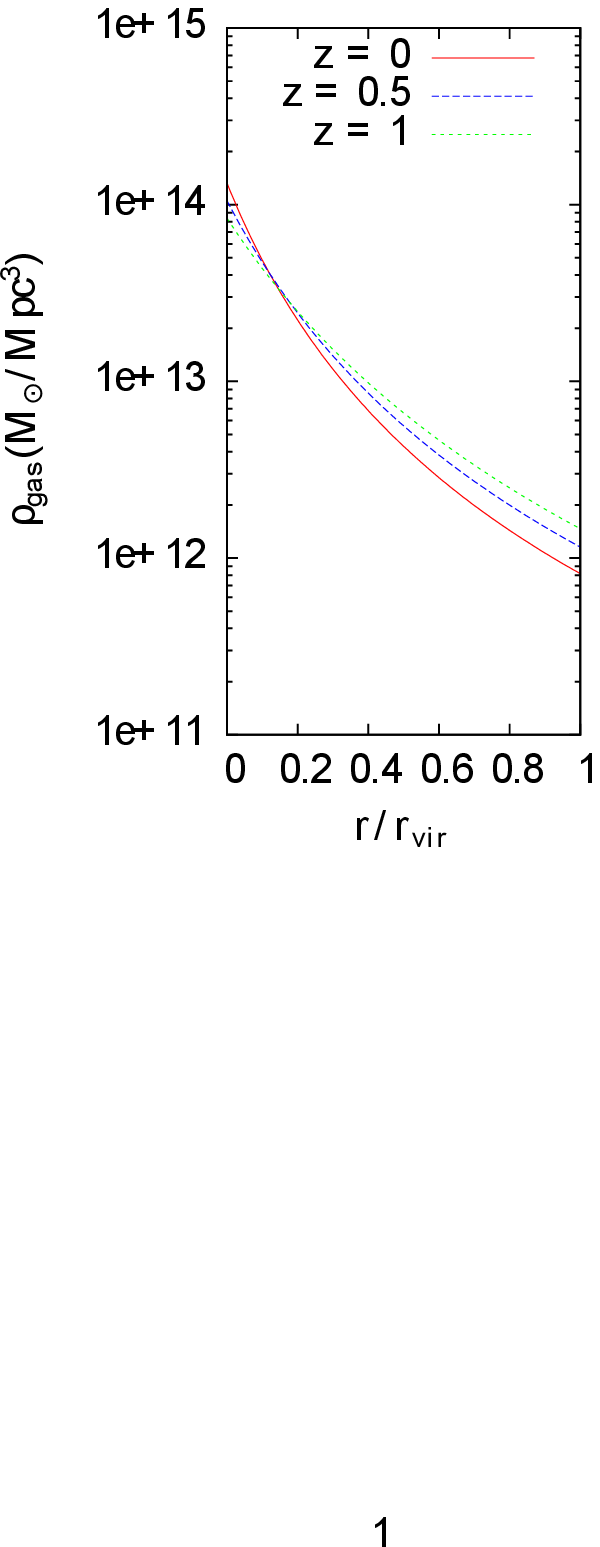}}
\end{subfigure}
\caption[Gas density profile]{The gas density profile for a halo of mass $10^{12}M_\odot$(left image) and $10^{15}M_\odot$(right image) varying with redshift.}
\end{figure}
\newpage
\begin{figure}[h!!!!!]
\centering
\hspace{0.35cm}
\begin{subfigure}[c]{0.47\textwidth}
\centering
\scalebox{0.95}{\includegraphics[trim={0 7cm 0 0},clip]{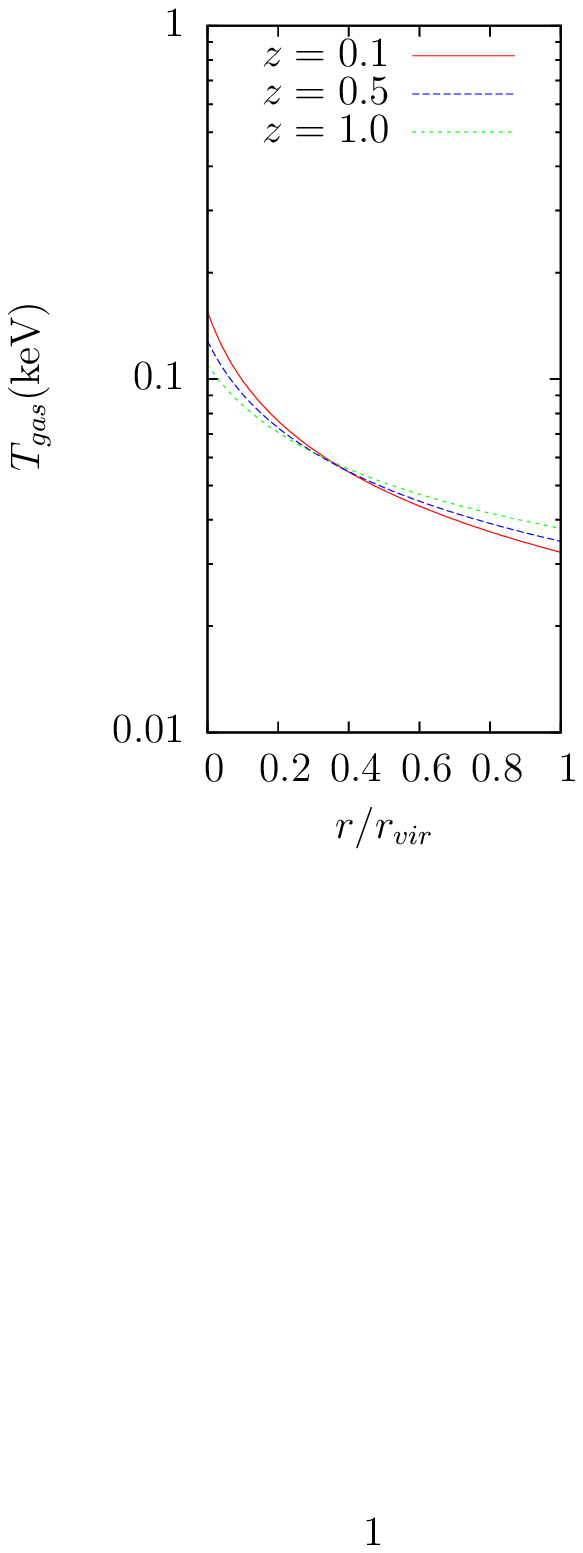}}
\end{subfigure}
\hspace{0.1cm}
\begin{subfigure}[c]{0.47\textwidth}
\centering
\scalebox{0.95}{\includegraphics[trim={0 7cm 0 0},clip]{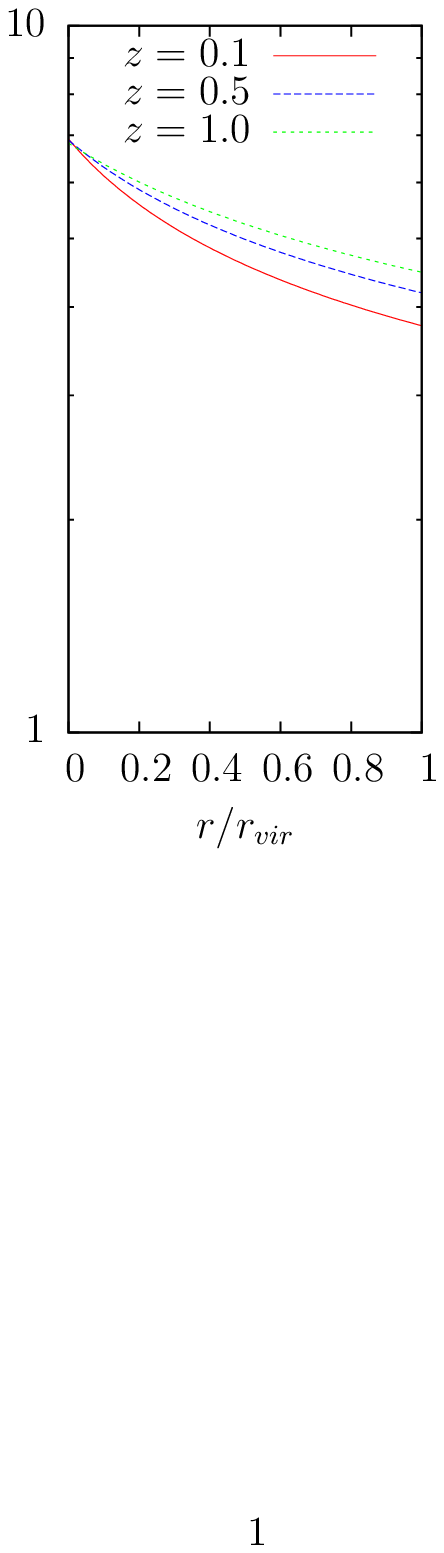}}
\end{subfigure}
\caption[Gas temperature profile]{The gas temperature profile for a halo of mass $10^{12}M_\odot$(left image) and $10^{15}M_\odot$(right image) varying with redshift.}
\vspace{1cm}
\begin{subfigure}[c]{0.47\textwidth}
\centering
\scalebox{0.95}{\includegraphics[trim={0 7cm 0 0},clip]{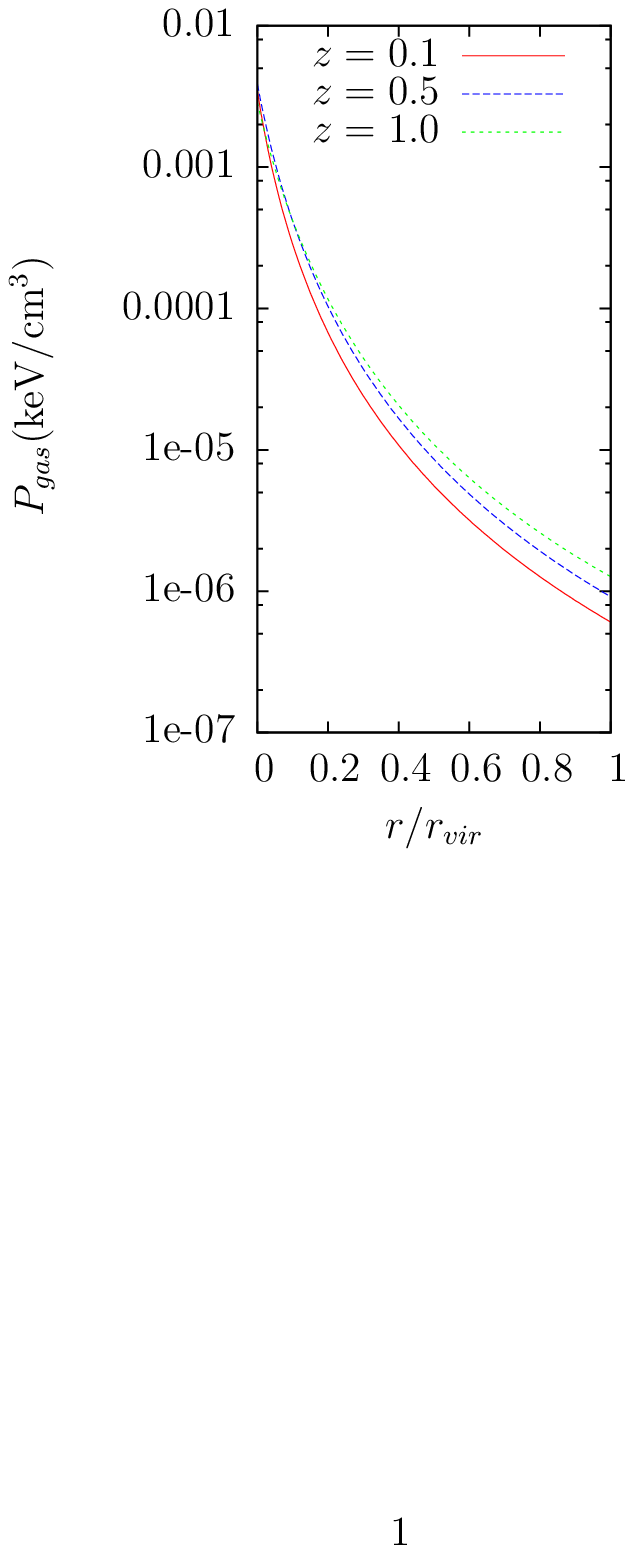}}
\end{subfigure}
\begin{subfigure}[c]{0.47\textwidth}
\centering
\scalebox{0.95}{\includegraphics[trim={0 7cm 0 0},clip]{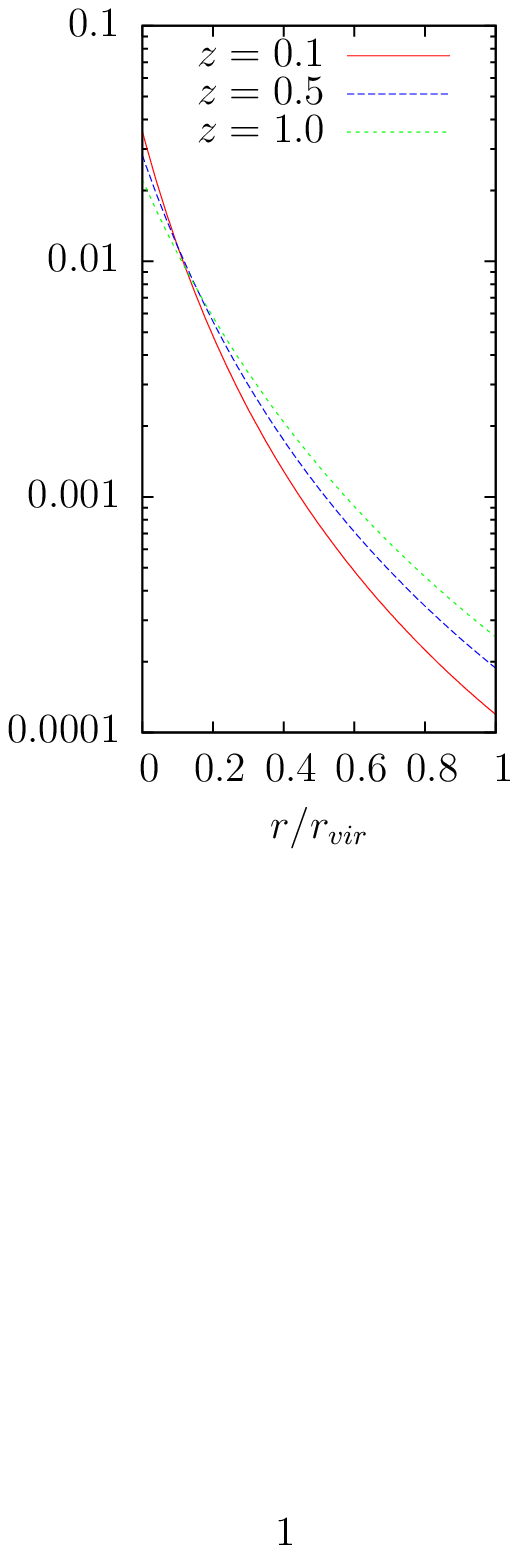}}
\end{subfigure}
\caption[Gas pressure profile]{The gas pressure profile for a halo of mass $10^{12}M_\odot$(left image) and $10^{15}M_\odot$(right image) varying with redshift.}
\end{figure}

\newpage

\section{Limitations of the Model}
The above model is only a first approximation of the gas in a dark matter halo. It does not take into account star-formation which will dominate in the central parts of the halo. The cooled down gas in the star-forming region will not have significant thermal energy to cause appreciable SZ effect (e.g., Shaw et al. 2010, Bode et al. 2009, Ostriker et al. 2005). In addition the gas can be heated by non-gravitational processes such as feedback from supernovae and halo mergers (e.g., Bode et al. 2009, Ostriker et al. 2005). Also there can be a significant non-thermal pressure support due to random gas motions and turbulence in the intracluster medium (Shaw et al. 2010). As the total pressure support, thermal and non-thermal, will balance the gravitational attraction of the dark matter in the halo, this will alter the thermal component of the gas pressure.

\chapter{Sunyaev-Zel'dovich Distortion from Host Halo Gas}
In this chapter I will use the model developed in the previous chapter to calculate the total SZ distortion due to intracluster gas in a halo.

\section{3D Compton y-Parameter Profile}
As shown in Chapter-1 Eq.\ 1.8, the SZ distortion along a line of sight from a halo is proportional to the Compton y-parameter defined as
\begin{align}
y &= \frac{\sigma_T}{m_e c^2}\int_l n_e(r)\ k_bT_e(r)\ dl \nonumber \\
&=\frac{\sigma_T}{m_e c^2}\int_l P_e(r)\ dl
\end{align}
where $\sigma_T$ is the Thompson scattering cross-section and $m_e c^2$ is the electron rest mass energy. $n_e(r)$ is the electron number density, $T_e(r)= T_{gas}(r)$ is the gas temperature in the halo and $P_e(r)$ is the thermal pressure of the electrons, each at a distance $r$ from the halo centre. The integral is over a line of light to the halo.

We define the 3D Compton y-parameter profile of a halo as
\begin{equation}
y_{3D}(r)=\frac{\sigma_T}{m_e c^2}\ P_e(r)
\end{equation}
Now, the electron gas pressure
\begin{align}
P_e(r) &= n_e(r)\ k_b T_{gas}(r) \nonumber 
\end{align}
Since the gas consists of hydrogen and helium, 
\begin{align}
n_e &= n_H +2n_{He} \nonumber \\
&= \frac{X\rho_{gas}}{m_p}+\frac{(2-2X)\rho_{gas}}{4m_p} \nonumber \\
&= \frac{\rho_{gas}}{4m_p}(2X+2) \nonumber
\end{align}
where X is the hydrogen mass fraction taken to be 0.75. Hence,
\begin{align}
 P_e(r) &= \frac{\rho_{gas}(2X+2)}{4m_p}\ k_b T_{gas}(r)
\end{align}
Using the expression for $P_{gas}(r)$ as in Eq.\ 2.32, I obtain
\begin{align}
P_e(r) &= \frac{2X+2}{5X+3}\ P_{gas}(r)
\end{align}
Thus, the 3D Compton y-parameter profile is given as
\begin{align}
y_{3D}(r) &= \frac{\sigma_T}{m_e c^2}\ \left(\frac{2X+2}{5X+3}\right)\ P_{gas}(r)
\end{align}

\begin{figure}[h!!!]
\centering
\begin{subfigure}[c]{0.3\textwidth}
\centering
\scalebox{0.75}{\includegraphics[trim={0 7cm 0 0},clip]{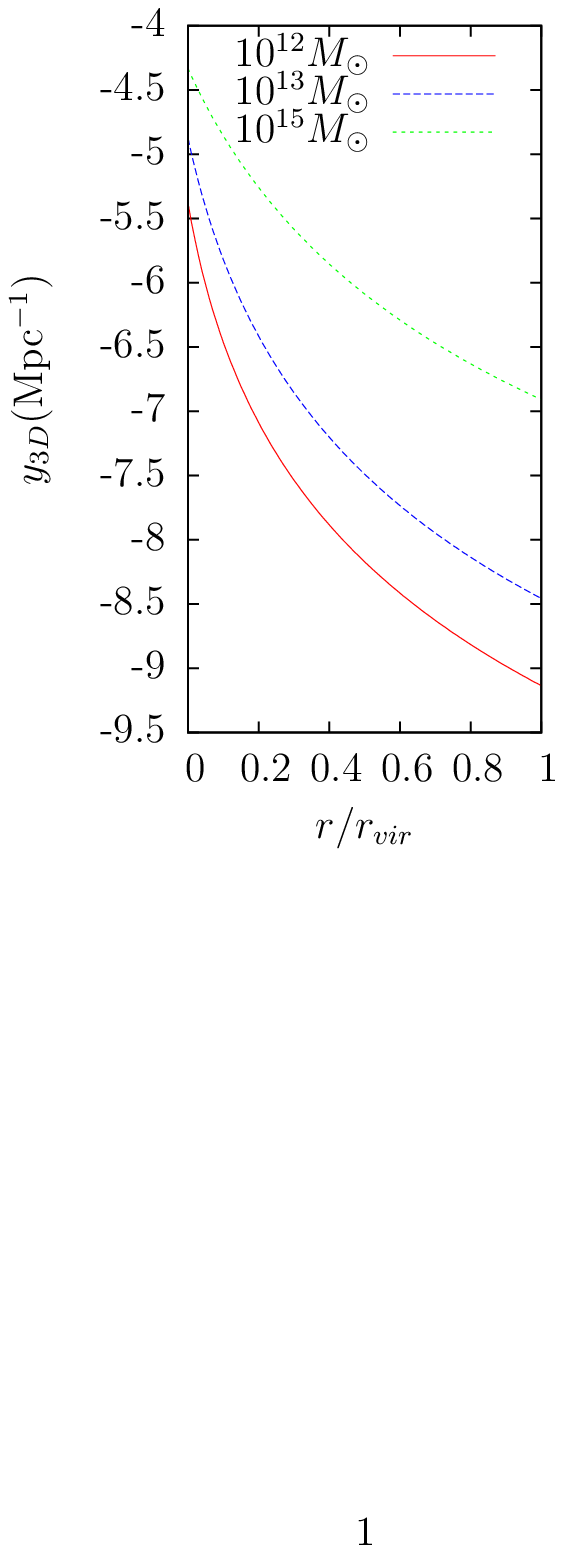}}
\end{subfigure}
\begin{subfigure}[c]{0.3\textwidth}
\centering
\scalebox{0.75}{\includegraphics[trim={0 7cm 0 0},clip]{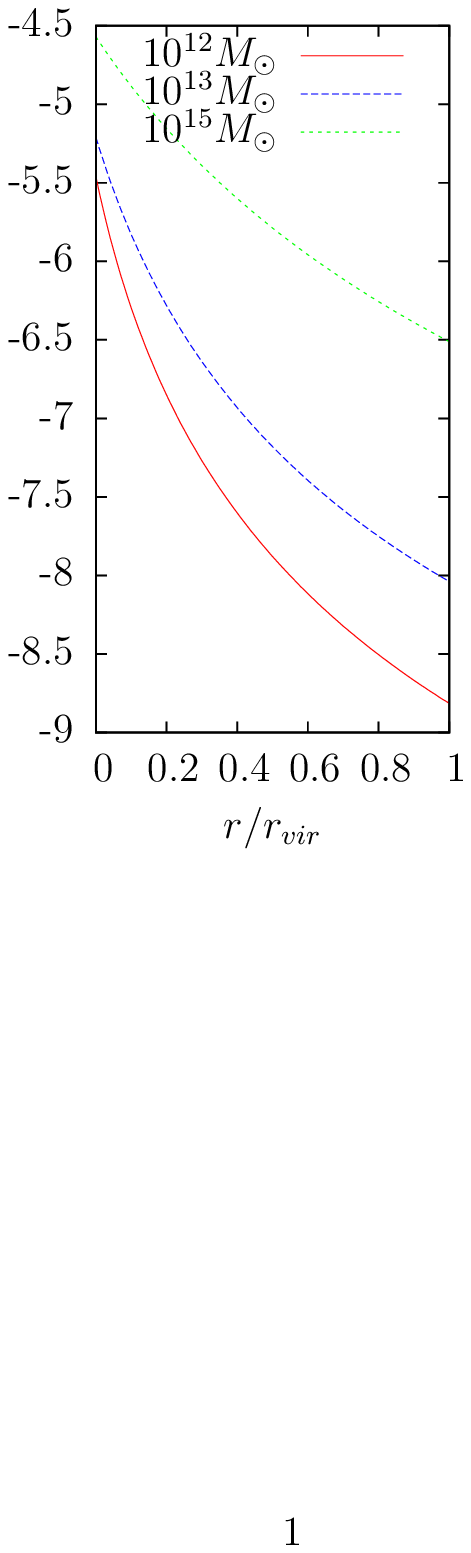}}
\end{subfigure}
\begin{subfigure}[c]{0.3\textwidth}
\centering
\scalebox{0.75}{\includegraphics[trim={0 7cm 0 0},clip]{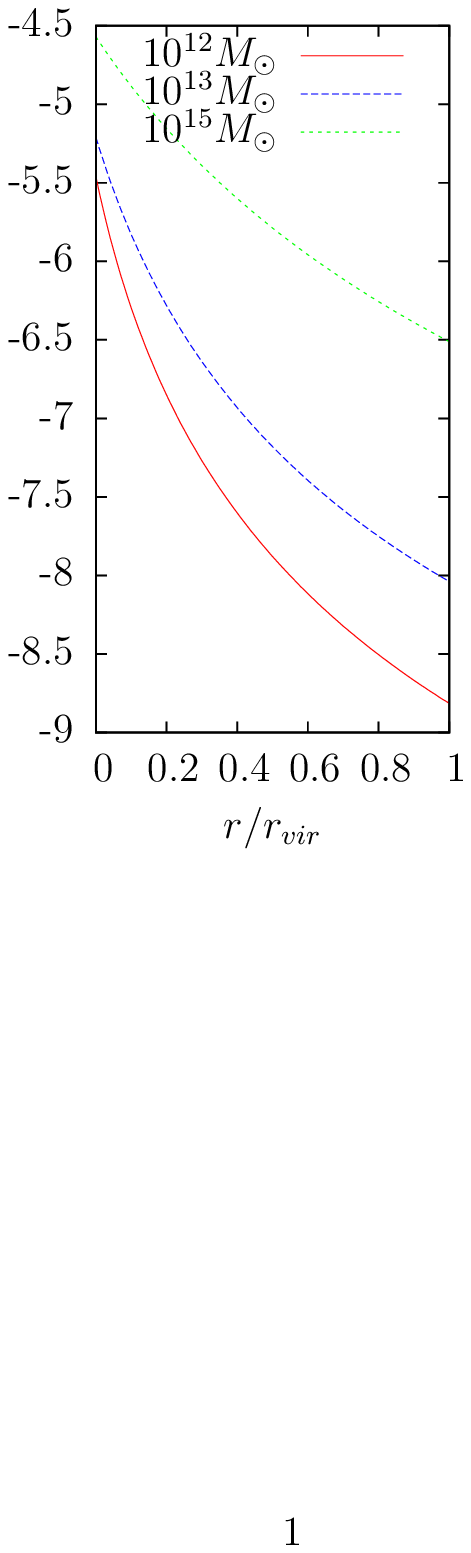}}
\end{subfigure}
\caption[3D Compton y-parameter profile]{The 3D Compton y-parameter profile for a halo of given mass. The left panel is at redshift 0, the middle one at 0.5 and the right panel at 1.0. The profiles vary little with redshift and at a given reshift is proportional to the halo mass.}
\end{figure}

\section{Total SZ Distortion from a Halo}
The total SZ distortion from a halo can be obtained by integrating the line of sight SZ distortion over the solid angle subtended by the halo at the observer ($z=0$) as explained in Chapter 1. Thus the integrated Compton y-parameter from a halo is given by (e.g., Calstrom et al. 2002)
\begin{align}
Y &= \int_\Omega y\ d\Omega \\
  &= \int_\Omega \int_l \frac{\sigma_T}{m_e c^2}\ P_e(r)\  dl\ d\Omega \nonumber \\
  &=\frac{1}{D^2_A}\int_V y_{3D}(r)\ dV 
\end{align}
where the last integral is over the total volume of the dark matter halo. $d\Omega = dA/D^2_A$, where $dA$ is a differential area element on the visible disk of the halo and $D_A$ is its angular diameter distance.

Using Eqns.\ 2.2, 2.33, 3.2, 3.7 and the definition of concentration parameter $c$, the integrated Compton y-parameter, $Y$ can be calculated as 
\begin{align}
Y &=\frac{1}{D^2_A}\int_V y_{3D}(r)\ dV \nonumber \\ 
  &= \frac{1}{D^2_A}\int_V \frac{\sigma_T}{m_e c^2}\ \left(\frac{2X+2}{5X+3}\right)\ P_{gas}(r) dV  \nonumber \\
  &= \frac{1}{D^2_A}\ \frac{\sigma_T}{m_e c^2}\ \frac{2X+2}{5X+3}\ \int_V P_{gas}(r) dV \nonumber \\
  &= \frac{1}{D^2_A}\ \frac{\sigma_T}{m_e c^2}\ \frac{2X+2}{4 m_p}\ \int_V \rho_{gas}(0)\ k_b T_{gas}(0)\ y^{\gamma}_{gas}(r/r_s) dV \nonumber \\
  &=\frac{1}{D^2_A}\ \frac{\sigma_T}{m_e c^2}\ \frac{2X+2}{4 m_p}\ \rho_{gas}(0)\ k_b T_{gas}(0)\ \frac{4\pi r^3_{vir}}{c^3} \int_0^c y^{\gamma}_{gas}(x)x^2 dx
\end{align}  
As $\rho_{gas}(0)$, $T_{gas}(0)$, $y_{gas}$ and $\gamma$ are functions of halo mass and redshift only the integrated Compton y-parameter from the halo ICM, $Y=f(M,z)$. 

Using Eq.\ 3.8, I calculate the integrated Compton y-parameter, Y(M,z) for haloes in the mass range $10^{12}M_\odot\ {\rm to}\ 10^{15}M_\odot$ and in the redshift intervel $z=0\ {\rm to}\ 1$. By convention Y(M,z) is expressed in square arcminutes. (Planck Collaboration XI 2013)

\newpage
\begin{figure}[h!!!]
\input{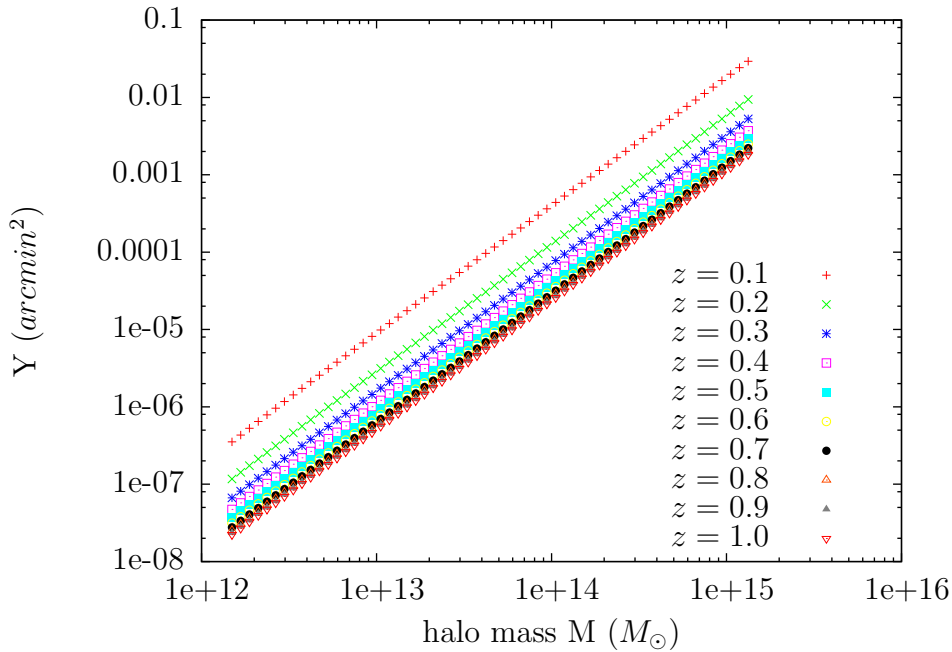}
\caption[Y-M relation at different redshifts]{Y-M relation calculated using Eq.\ 3.8 due to virialised halo gas at different redshifts from z= 0 to z=1. At a given redshift the Y-M relation appears to be a straight line in log-log plot implying a power law scaling of Y with respect to halo mass}
\end{figure}

\section{Rescaling the SZ Signal}
Due to differences in angular-diameter distance, in order to compare the integrated Compton y-parameter at different redshifts in my work and to those of Ruan et al. (2015), the SZ signal is rescaled to a common $D_A\ =\ {\rm 500\ Mpc}$. The expected redshift dependence of the signal due to virialised halo gas is also divided out to get the rescaled Compton y-parameter $Y^{re}$ following the prescription of Planck Collaboration XI (2013). This is also expressed in square arcminutes as
\begin{equation}
Y^{re} = Y\ E^{-2/3}(z)\ \left(\frac{D_A}{500\ {\rm Mpc}}\right)^2
\end{equation}
where $E^2(z)= \Omega^0_m(1+z)^3+\Omega_\Lambda$ 
\newpage
\begin{figure}[h!!!!!!]
\input{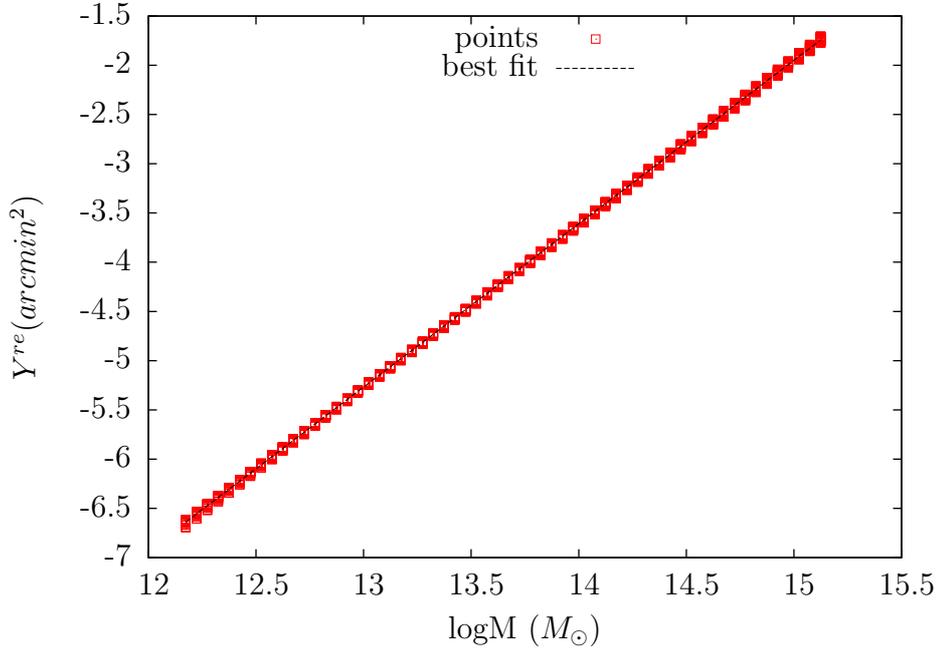}
\caption[$Y^{re}-M$ relation]{$Y^{re}-M$ relation. The log-log plot is a straight line independent of redshift implying a self-similar power law scaling of $Y^{re}$ with the halo mass M. The power-law exponent is obtained as $a=1.66$}
\end{figure}

In Fig.\ 3.3, I plot $Y^{re}$ due to the virialised halo gas. The profiles for different redshifts merge to give a single straight line implying the $Y^{re}-M$ relation to be independent of redshift as expected. For a halo of fixed mass, the rescaled y-parameter is the same for all redshifts.

A linear fit to the data gives the power-law slope and the normalization as a $= 1.66$ and b $= 10^{-26.84}$ respectively. 
Thus we have a self similar relation between $Y_{re}$ and M given by
\begin{align}
Y^{re} &= b\ M^{a} \\
or,\ \frac{Y^{re}}{Y^{re}_s} &= \left(\frac{M}{M_s}\right)^{a}
\end{align}
where $Y^{re}_s$ is the rescaled Compton parameter for a characteristic halo mass $M_{s}$. The characteristic halo mass is chosen as 
$M_{s}= 3.0 \times 10^{14} M_\odot$ following Planck Collaboration XI (2013).

From Eq.\ 3.10 
\begin{align}
b=\frac{Y^{re}_s}{M_s^{a}} &= 10^{-26.84} \nonumber \\
Y^{re}_s &= 10^{-26.84}\times M_s^{a}  \nonumber \\
&= 1.55\times 10^{-3}
\end{align}

Thus the self similar $Y^{re}-M$ relation is obtained as
\begin{align}
Y^{re} &= 1.55\times 10^{-3} \left(\frac{M}{3\times 10^{14}M_\odot}\right)^{1.66}
\end{align}

The $Y^{re}-M$ relation obtained above is compared to that observationally estimated by Planck Collaboration XI (2013). Planck Collaboration XI (2013) in their work stacked the Planck SZ maps in order to estimate the mean SZ signal from locally bright galaxies (LBGs) in a series of stellar mass bins. They used mock galaxy catalogues based on the Millennium Simulation and tuned to fit the observed abundance and clustering of SDSS galaxies to establish the relation between stellar and halo mass and found that the $Y^{re}-M$ data agrees well with a single power law within its statistical uncertainties given by

\begin{equation}
Y^{re} = (0.73 \pm 0.07)\times 10^{-3} \left(\frac{M}{3\times 10^{14}M_{\odot}}\right)^{5/3}.
\end{equation}

I note that the observational results support the self similar scaling relation, which justifies my theoretical assumption and the model of Komatsu \& Seljack. 

The normalisation factor in my model is nearly double to that of the Planck estimation. This is because the Planck team analyses the halo upto a radius $R_{500}$, which defines a sphere inside which the average dark matter density is 500 times the mean matter density at that epoch. However in my model the contribution to SZ signal is taken from gas upto the virial radius of the halo, which is defined in terms of the overdensity parameter with respect to the critical density in Eq.\ 2.7. Hence, in my model integration is done over a larger volume of gas which results in a difference between the normalisaions. 

\chapter{Average Sunyaev-Zel'dovich Signal from Quasar Hosts}
In the previous chapter the rescaled SZ signal due to virialised gas from a single halo of mass M was obtained given by Eq.\ 3.13. It was found to be independent of redshift. In this chapter, I will model the average SZ signal due to virialised gas of quasar hosting haloes at a given redshift. The average signal has a redshift dependence because the number of quasar hosting haloes will depend on redshift both due to the redshift dependence of the halo mass functions and the redshift dependence of the quasar hod parameters (e.g., Sheth \& Torman 1999; Jenkins et al.\ 2001; Chatterjee et al. 2012). 

Thus to find the average signal, I will first model the number of quasar hosting haloes at a given redshift using the Halo Mass Function (Sheth \& Torman\ 1999) and the Halo Occupation Distribution of quasars (Richardson et al.\ 2012, Chatterjee et al.\ 2012).

\section{Halo Mass Function}
The Halo Mass Function (HMF), $\frac{dn}{dM}(M,z)$ is defined as the number of dark matter haloes of mass $M$ per unit mass interval, per unit comoving volume at a given redshift $z$. Thus, it is the comoving number density of dark matter haloes per unit mass interval. The redshift evolution of the mass function is dependent on the cosmology. A halo is defined as an approximately spherical region of dark matter with average density much greater, typically 200 times the mean matter density at that epoch. In the standard model of cosmology with cold dark matter, small and less massive haloes form first which then merge together to form more maasive haloes. Thus the number of massive haloes decreases with increasing redshift (earlier epoch). Assuming a spherical or an ellipsodial collapse model for dark matter particles, HMF can be analytically calculated following the prescriptions of Press \& Schechter (1974) and Sheth \& Torman (1999) respectively. It can also be obtained from high resolution N-body simulations. (e.g., Jenkins et al. 2001; Tinker et al. 2008; Klypin et al. 1999; Moore et al. 1999; Davis et al. 1985; Lacy \& Cole 1994; Reed et al. 2003; Warren et al. 2006; Peacock et al. 2007)

In my work, I use the Sheth \& Torman (1999) mass function. Sheth \& Tormen (1999) assumed an ellipsoidal collapse model of dark matter particles rather than a spherical collapse model to account for the discrepancy between the Press \& Schechter (1974) model and N-body simulations. In the original spherical model, a region collapses if the initial density within it exceeds a threshold value, $d_{sc}$ . This value is independent of the initial size of the region and since the mass of the collapsed object is related to its initial size, $d_{sc}$ is independent of final mass. In the ellipsoidal model, the collapse of a region depends on the surrounding shear field, as well as on its initial overdensity. Since for Gaussian random fields, the distribution of these quantities depend on the size of the region considered, there is a relation between the threshold density value required for collapse and the mass of the final object. The Sheth \& Tormen (1999) mass function is analytically given by 

\begin{align}
\frac{dn}{d{\rm ln}\ M} &= \frac{\rho_{m}}{M} f(\nu) \frac{d \nu}{d{\rm ln}\ M} \\
\nu f(\nu ) &= 2A \left(1+\frac{1}{\nu^{2q}}\right) \left(\frac{\nu^{2}}{2 \pi}\right)^{1/2} exp\left(-\nu^{2}/2 \right)
\end{align}
Here q=0.3, A=0.3222 and $\nu =\frac{\delta_{sc}(z)}{\sigma_{m}}$ is the ratio of the critical overdensity required for collapse in the spherical model to the rms density fluctuation on a scale equivalent to the size of the collapsing mass M (Sheth \& Tormen, 1999)

To obtain the halo mass function, I use the numerically generated data files from HMFcalc. HMFcalc is a web application for calculating the HMF (Murray, Power \& Robotham, 2013). By choosing the cosmological parameters, the desired redshift, overdensity parameter, range of halo masses and the bin size for halo mass as run parameters the data files can be generated for any desired HMF. I generated the data files in the redshift range $0$ to $1$ and halo mass range $10^{12} M_{\odot}$ to $10^{15} M_{\odot}$ with mass bin size $10^{0.05} M_{\odot}$. Shown in Fig\ 4.1 is a log-log plot of the mass function at different redshifts.

\begin{figure}[h!!!]
\input{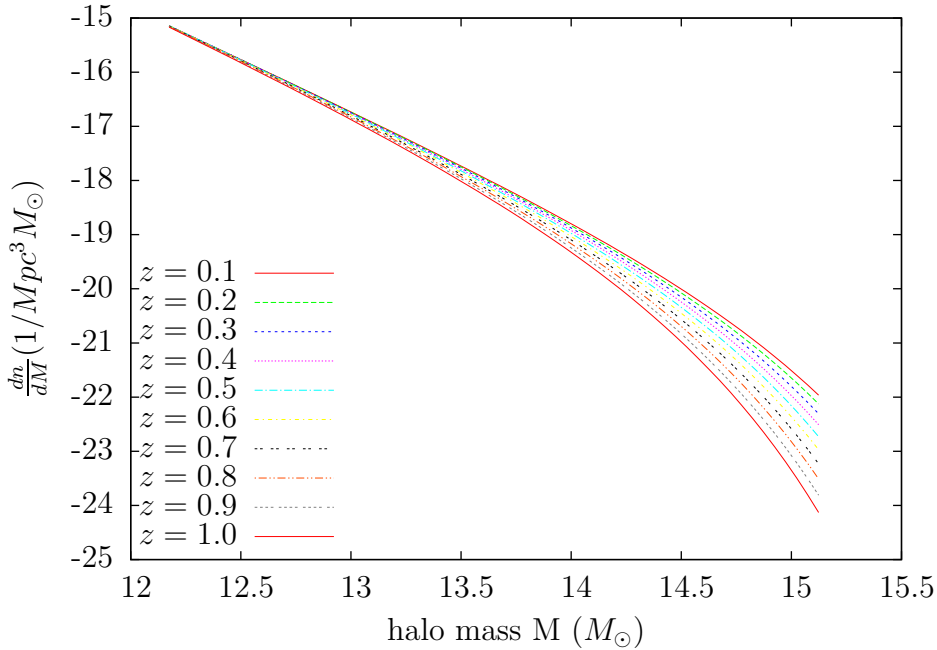}
\caption[Sheth \& Tormen Halo Mass Function (1999)]{Sheth \& Tormen (1999) Halo Mass Function at different redshifts obtained with HMFcalc data files}
\end{figure}

\section{Halo Occupation Distribution of Quasars}
The growth of black-holes or AGNs in dark matter haloes is governed by the properties of the dark matter particles and the gas dynamics in the host halo. The clustering properties of AGNs can be studied using semi-analytic techniques (e.g., Shankar et al. 2010; Bonoli et al. 2009; Lidz et al. 2006) while the environment dependence of accretion can be modelled through hydrodynamic simulations of galaxy formation with black hole growth (e.g., Thacker et al. 2009; Degraf et al. 2011). Such simulations can be used to effectively study the co-evolution of AGNs with dark matter haloes and host galaxies. However due to the complex parameter spaces of these models and the time it takes to run simulations make difficult to work with.

The Halo Occupation Distribution (HOD) (e.g., Ma \& Fry 2000; Seljak 2000; Berlind \& Weinberg 2002) is an alternative approach where the entire astrophysics determining the connection between AGNs and their host haloes is encoded in an analytic probabilty distribution function characterised by the mass of the host halo. It can be used to study the clustering properties of AGNs. The HOD for AGNs is defined by the conditional probability $P\ (N\ |M )$ that a halo of mass $M$ contains $N$ AGNs of a given type (generally based on luminosity), along with their spatial and velocity distribution inside the halo. This formalism distinguishes between the background cosmology and AGN evolution models as the cosmological parameters are encoded in the distribution of haloes given by the HMF, where as the relation between halo mass and AGN is fully described by the probability distribution $P\ (N\ |M )$ (Chatterjee et al. 2012). 

The mean occupation function for AGNs of a given type, $\langle N(M) \rangle$, is the average number of AGNs of that type in a halo of mass M. Assuming $P\ (N\ |M )$ to be normalised, the mean occupation fraction is given by

\begin{equation}
 \langle N(M) \rangle = \Sigma_0^{N_{max}}\ N\ P\ (N\ |M)
\end{equation}
 
It can be represented as the sum of two components- central mean occupation fraction, $\langle N^{c}(M) \rangle$ and satellite mean occupation fraction, $\langle N^{s}(M) \rangle$. (e.g., Kravtsov et al.\  2004; Zheng et al.\  2005; Chatterjee et al. 2012)   

I use the Chatterjee et al. (2012) HOD model in my work. Chatterjee et al. (2012) did a study of low luminosity AGNs in cosmological hydrodynamic simulations to obtain their mean occupation function. In this parameterization the central AGN was chosen as the most massive black hole within $R_{200}$ and the rest of the black holes within $R_{200}$ were taken as satellite AGNs. According to this model, the mean occupation function of central AGNs is given by a softened step function and for the satellite AGNs it is a rolling off power law.
\begin{align}
\langle N^{c}(M) \rangle &= \frac{1}{2}\left[1+erf\left(\frac{{\rm log}\ M - {\rm log}\ M_{min}}{\sigma_{{\rm log}\ M}}\right)\right] \\
\langle N^{s}(M) \rangle &= \left(\frac{M}{M_{1}}\right)^{\alpha}\ exp\left(-\frac{M_{cut}}{M}\right) \\
\langle N(M) \rangle &= N^{c}(M) \rangle + \langle N^{s}(M) \rangle
\end{align}

The model has five free parameters namely $M_{min}$, $\sigma_{{\rm log}\ M}$, $M_{1}$, $\alpha$ and $M_{cut}$. $M_{min}$ is the approximate mass scale at which on average half of the haloes have one quasar at their centre; $\sigma_{{\rm log}\ M}$ is
the characteristic transition width of the softened step function; $M_1$ is the approximate mass scale at which haloes on average have one satellite quasar; $\alpha$ is the power law exponent of the mean satellite occupation function; and $M_{cut}$ is the mass scale below which the satellite mean occupation decays exponentially. 

This model although obtained from study of low-luminosity AGNs can be extended to quasars (most luminous AGNs) as Richardson et al. (2012) showed that it explains the observed 2 pt correlation functions of quasars. Richardson et al. (2012) determined the free parameters of the model by fitting the 2-point correlation function calculated from this model with observed data from SDSS quasars at $z= 1$. The best fit values for the same are given in Table\ 4.1. For the current work I used these values to compute the quasar HOD.

\begin{table}[h!!!]
\centering
\begin{tabular}{|c|c|}
\hline
Free parameters & Best fit value \\
\hline
$M_{min}$ & $10^{16.46}$ \\ 
$\sigma_{{\rm log}\ M}$ & 1.667 \\ 
$M_{1}$ & $10^{12.47}$ \\ 
$\alpha$ & 0.6158 \\
$M_{cut}$ & $10^{15.28}$ \\ \hline
\end{tabular}
\caption[Best fit values of HOD model from Richardson et al. (2012)]{Best fit values of the free parameters for the quasar HOD in the Chatterjee et al. (2012) model as determined by Richardson et al. (2012)}
\end{table}

Using these parameters, I numerically calculate the central, sattelite and total mean occupation of quasars as a function of halo mass as shown in fig.\ 4.2
\begin{figure}[h!!!!!!!!!!]
\input{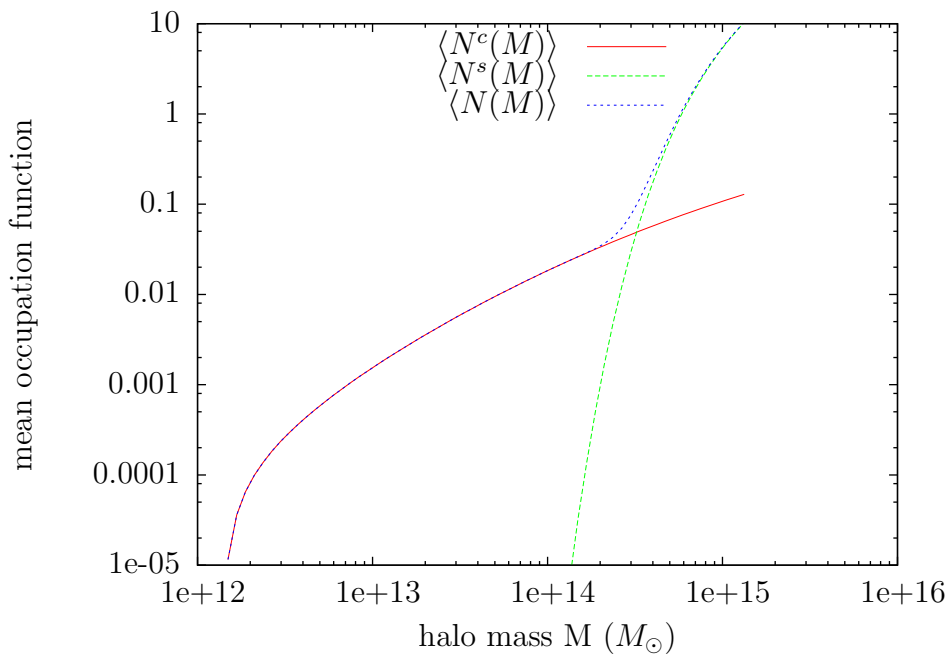}
\caption[Mean Occupation Function of Quasars]{The central, sattelite and total mean occupation of quasars in host haloes as a function of halo mass}
\end{figure}

Chatterjee et al. (2012) showed that the HOD parameters evolve with redshift for low luminosity AGNs. However due to lack of measured redshift evolution of quasar HODs, I shall assume the mean occupation function of quasars to be independent of redshift specified at all redshifts by the same parameters as those at $z=1$.

\section{Number Density of Quasar hosts}
The HMF gives the total number of haloes per unit comoving volume at a given redsift. However all haloes will not be hosting quasars. The fraction of total haloes of mass $M$ per unit mass intervel per unit comoving volume that host quasars, is given by the mean occupation function of quasars, $\langle N(M) \rangle$. Thus the number of quasar hosting haloes of mass $M$ at redshift $z$, $N_{q}(M,z)$ per unit mass per unit comoving volume can be modelled by
\begin{equation}
N_{q}(M,z)= \langle N(M) \rangle\ \frac{dn}{dM}(M,z)
\end{equation}
In this model the redshift dependence in the number of quasar hosts comes from the redshift dependence of the halo mass function. 

The number of quasar hosts per unit volume per unit mass intervel is calculated following Eq.\ 4.7 and plotted as a function of halo mass at $z=1.0$.

\begin{figure}[h!!!!]
\input{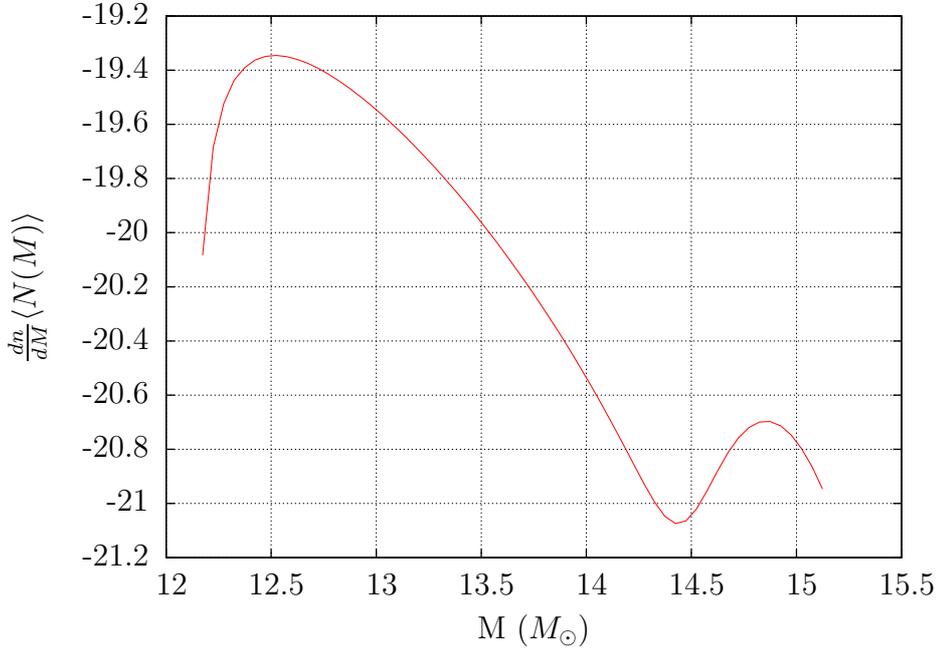}
\caption[Number of quasar hosting haloes]{Number of quasar hosting haloes per unit mass per unit comoving volume at $z = 1.0$. The first peak at $M=10^{12.5}M_{\odot}$ represents the maximum in the number density of central quasar occupying haloes and the second peak at $M=10^{14}M_{\odot}$ represents the maximum in the number density of satellite qusar occupying haloes.}
\end{figure}

$N_{q}$ has a local maximum at around $M=10^{12.5}M_{\odot}$. At this halo mass, the satellite mean occupation is negligable and the total mean occupation is entirely due to central quasars. Thus the number of haloes hosting central quasars peaks at around halo mass $M=10^{12.5}M_{\odot}$. Another peak in $N_{q}$ occurs around $M=10^{14.8}M_{\odot}$. This occurs because for haloes having mass $M=10^{14}M_{\odot}$ and above the satellite mean occupation of quasars increases rapidly and overshoots the central population. Thus this mass peak represents the maxima in the number of haloes hosting satellite quasars.

\section{Average SZ signal}

The average SZ signal at a given redshift due to virialised halo gas is proportional to the average rescaled Compton Y-parameter and is given by
\begin{align}
\langle Y^{re}(z) \rangle &= \frac{\int_{M_1}^{M_2} Y_{re}(M)\ N_{q}(M)\ dM\ dV}{\int_{M_1}^{M_2} N_{q}(M)\ dM\ dV}
\end{align} 
where dV is the comoving volume between redshift $ z\ to\ z+dz$. Using Eq.\ 4.6 
\begin{align}
\langle Y^{re}(z) \rangle &= \frac{\int_{M_1}^{M_2} Y_{re}(M)\ \langle N(M) \rangle\ \frac{dn}{dM}\ dM}{ \int_{M_1}^{M_2} \ \langle N(M) \rangle\ \frac{dn}{dM}\ dM} 
\end{align}

Using the self similar fitting function obtained for $Y^{re}$ in Eq.\ 3.13, the integral in Eq.\ 4.9 is computed for redshifts varying from 0 to 1. The limits of integration are taken as $M_{1}= 10^{12}M_{\odot}$ and $M_2=10^{15}M{\odot}$ as it is seen in the previous section that most of the quasar hosting haloes are in this mass range. The calculated values are tabulated in Table 4.2. With increasing redshift from 0 to 1, the average rescaled Compton y-parameter decreases from $\sim (10^{-4}$ to $10^{-5})$

As the quasar HOD is assumed to be redshift independent, the redshift evolution of the number of quasar hosting haloes follows that of the HMF. At lower redshifts, there are significant number of high mass haloes ($10^{14} M_{\odot}-10^{15} M_{\odot}$) as compared to high redshifts around 1. From Eq.\ 3.15, $Y_{re}$ from them is of the order of $10^{-4}-10^{-3} {\rm arcmin}^{2}$. With increasing redshift, the number of high mass haloes decreases and the majority of haloes are in the mass range $10^{12} M_{\odot}-10^{13} M_{\odot}$. $Y_{re}$ from these haloes is of the order of $10^{-6}-10^{-5} {\rm arcmin}^{2}$. This might explain the decreasing trend obtained in $\langle Y^{re} \rangle$ with increasing redshift.

\begin{table}[h!!!!!]
\centering
\begin{tabular}{|c|c|}
\hline
redshift z & average Compton Y ($\langle Y^{re}(z) \rangle$) \\
\hline
0.1 & 1.76\ E\ -04 \\
0.2 & 1.38\ E\ -04 \\
0.3 & 1.06\ E\ -04 \\
0.4 & 8.01\ E\ -05 \\
0.5 & 6.00\ E\ -05 \\
0.6 & 4.46\ E\ -05 \\
0.7 & 3.32\ E\ -05 \\
0.8 & 2.50\ E\ -05 \\
0.9 & 1.91\ E\ -05 \\
1.0 & 1.50\ E\ -05 \\ \hline
\end{tabular}
\caption{The rescaled average Compton y-parameter at different redshifts}
\end{table}

\begin{figure}[h!!!!]
\input{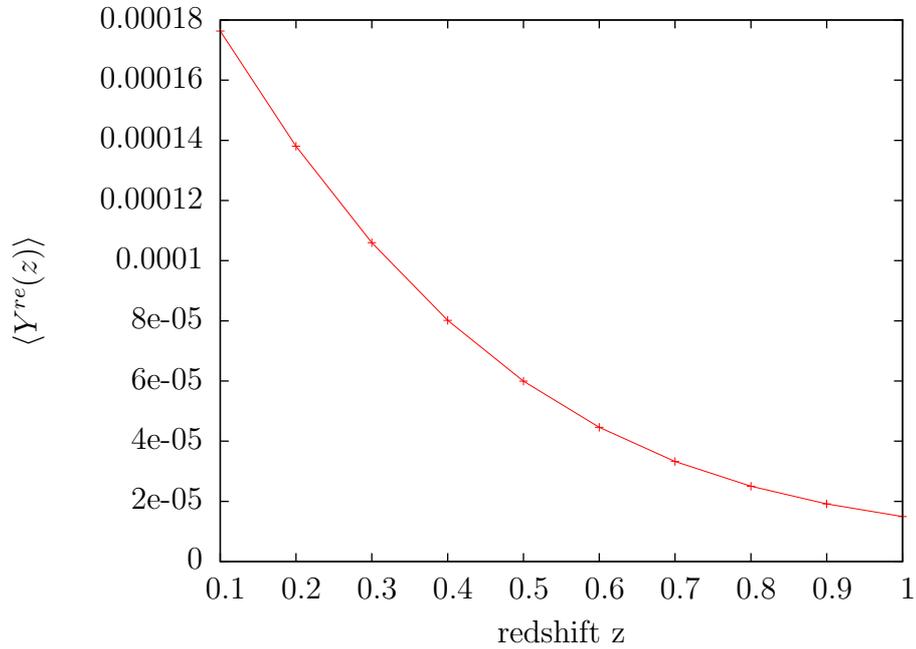}
\caption[Rescaled Compton y-parameter varying with redshift]{The rescaled average Compton y-parameter varying with redshfit. A decreasing trend is observed with increase in redshift from 0 to 1}
\end{figure}

\chapter{Discussion of Results}
Very recently, Ruan et al.\ (2015) claimed to have detected signals of quasar feedback in the TSZ Compton y-maps cross-correlated with the locations of 26,686 SDSS spectroscopic quasars. The integrated TSZ signal detected in the stacked Compton y-maps is likely to be originating from both interstellar gas of the host galaxy heated by quasar feedback as well as virialised halo gas in quasar hosts. 

To isolate the quasar feedback signal, two different stacks of quasar subsamples from the total SDSS sample was created, one at high redshift ($z \sim 2$) and the other at low redshift ($z \sim 1$). Ruan et al.\ (2015) assumed that the high-redshift subsample will not contain quasars lying in massive clusters and thus its TSZ signal will be purely due to feedback effects without any contamination from the host halo. From this signal, they derived the quasar feedback energetics. They further showed that the low redshift stacked signal is roughly consistent with that estimated from a combination of quasar feedback at low redshift with similar feedback parameters as that at high redshift plus a virialised halo gas component from quasar hosts. The host halo mass distribution of quasars was obtained from quasar clustring data by Shen et al.\ (2013).

Using my estimation for the average Compton Y-parameter, $\langle Y^{re}(z) \rangle$, due to the virialised gas of quasar hosts at both low and high redshifts, I will check if the virialised gas signal at high redshifts can at all be neglected compared to that at low redshifts as assumed by Ruan et al. (2015). I have already obtained trends in my work which show that $\langle Y^{re}(z) \rangle$  decreases with redshift. However the best fit values of the HOD model parameters as determined by Richardson et al. (2012) have errors. Using these errors, the error in $\langle Y^{re}(z) \rangle$ needs to be estimated. If the results are positive, then using my estimate for $\langle Y^{re}(z) \rangle$ at low redshift and the feedback parameters determined from the high redshift quasar subsample by Ruan et al. (2015), I will check whether the stacked SZ signal from the low redshift quasar subsample of Ruan et al. (2015) can be reproduced. Thus, my work will provide a theoretical route to estimate the significance of the results obtained by Ruan et al. (2015).  

It will be the first effort to clearly validate the thermal SZ effect from quasar feedback, a signal that has been predicted by different groups (e.g., Natarajan \& Sigurdsson 1999; Yamada et al. 1999; Lapi et al. 2003; Platania et al. 2002; Chatterjee \& Kowosky 2007) 

The gas profile used for calculating the Y-M relation may also be modified if needed taking into account star formation, non-thermal pressure and supernova feedback (Shaw et al. 2010) and the results compared with that obtained from the Komatsu \& Seljak model. Finally I will incorporate the redshift dependence of the quasar HOD in my work.

\end{document}